\documentclass[aps,prd,reprint,amsmath,amssymb,superscriptaddress,floatfix]{revtex4}
\usepackage{graphicx}
\usepackage{verbatim} 
\usepackage{amsfonts}
\usepackage{amssymb}
\usepackage{rotating}
\usepackage{booktabs}
\usepackage{xcolor}
\usepackage{soul}
\usepackage{color}
\usepackage{slashed}
\usepackage{multirow}
\usepackage{makecell}
\usepackage{epsf}
\usepackage{ulem}
\usepackage{cancel}
\usepackage{color,bm}
\usepackage[colorinlistoftodos]{todonotes}
\usepackage{diagbox}
\usepackage[colorlinks=true,citecolor=cyan,urlcolor=blue,bookmarks=true,bookmarks=true,bookmarksopen=true,bookmarksnumbered=true,bookmarksopenlevel=3]{hyperref}
\usepackage{subfigure}
\definecolor{airforceblue}{rgb}{0.36, 0.54, 0.66}
\definecolor{steelblue}{rgb}{0.27, 0.51, 0.71}
\definecolor{amber}{rgb}{1.0, 0.49, 0.0}

\newcommand{\s}{\slashed}

\begin{document}

\title{Scalar dark matter with $Z_3$ symmetry in Type-II seesaw}
\author{\textsc{XinXin Qi}}
\author{\textsc{Hao Sun}\footnote{Corresponding author: haosun@dlut.edu.cn}}
\affiliation{ Institute of Theoretical Physics, School of Physics, Dalian University of Technology, \\ No.2 Linggong Road, Dalian, Liaoning, 116024, People’s Republic of China }
\date{\today}

\begin{abstract}
We study a simple complex scalar singlet dark matter (DM) model with $Z_3$ symmetry in the framework of the type-II seesaw mechanism.
We use the model to explain the excess of electron-positron flux measured by AMS-02, DAMPE and Fermi-LAT collaborations,
which is encouraged by the decay of the triplets produced from dark matter annihilations in the Galactic halo.
We focus on the non-degenerate case in which the mass of DM is larger than that of the triplets' 
and deliberately alleviates the leptophilic properties of the DM, so that the semi-annihilation effects are enhanced.
With the guarantee of $Z_3$ symmetry, by fitting the antiproton spectrum observed in PAMELA and AMS experiments,
we find that the DM cubic terms and the couplings between DM and Higgs are strongly constrained,
leading to the semi-annihilation cross section fraction less than $3\%$ when DM mass is given at 3 TeV.

\vspace{0.5cm}
\end{abstract}
\maketitle
\setcounter{footnote}{0}

\section{Introduction}
\label{sec:intro}
  So far, there are at least two unsolved problems in particle physics: 
the nature of dark matter (DM) proved by astronomical evidence \cite{Zwicky:1933gu,Allen:2011zs,Rubin:1980zd,Frenk:2012ph,Refregier:2003ct,Hinshaw:2012aka}
and the origin of the neutrino mass revealed by the observation of neutrino oscillations \cite{Al2014Review}. For the dark matter, one of the most attractive candicate is the so-called weakly interacting massive particles (WIMPs), and the observed dark matter relic density can be generated by the Freeze-out mechanism.
Another paradigms of DM include Strongly Interacting Massive Particles (SIMPs) and Forbidden dark matter, 
where SIMPs are realized by (effective) five-point self-interactions \cite{Choi:2015bya,Hochberg:2014dra,Hochberg:2014kqa,Lee:2015gsa}, 
and the latter is the annihilation of hidden sector set opened at high temperature \cite{Choi:2015bya,Choi:2016tkj,DAgnolo:2015ujb}.
On the other hand, in order to understand the origin of neutrino mass theoretically, several mechanisms have been proposed,
such as type-I seesaw \cite{Minkowski:1977sc,1979Workshop,Mohapatra:1979ia}, by introducing heavy right-handed neutrinos,
type-II seesaw \cite{Magg:1980ut,Lazarides:1980nt,Ashery:1981tq,Ma:1998dx,Konetschny:1977bn,Schechter:1980gr,Cheng:1980qt,Bilenky:1980cx}, 
by adding $SU(2)$ Higgs triplet, and so on \cite{Ma1998Pathways,Bajc2006Seesaw,P2007Renormalizable,Perez:2007iw}.

In this paper, we consider the dark matter issue as well as neutrino mass issue in a common framework.
For the settlement of DM, we introduce a complex scalar singlet ($S$) with a discrete $Z_3$ symmetry, and its $Z_3$ charge is unity ($X_{S}=1$) 
while the SM particles' are zero $(X_{\rm SM}=0)$, the stability of DM can therefore be guaranteed by the $Z_3$ symmetry. Such type of the extended singlet scalar could be naturally embedded in the $SO(10)$ group on account of the same gauge and $B-L$ quantum number compared with SM fermions.
 On the other hand, discrete symmetry such as $Z_2$, $Z_4$ or $Z_N$ dark matter model have been presented in, for example, 
Refs.\cite{1978Pattern,Ma:2006km,Barbieri:2006dq,LopezHonorez:2006gr,Belanger:2014bga,Belanger:2012vp},
where the common feature of such models is that the discrete symmetry could be the remnant symmetry of some breaking $U(1)_X$ gauge group \cite{Martin:1992mq,Krauss:1988zc}. It is worth stressing that the semi-annihilation processes can arise if cubic terms exist among the dark matter or dark sector \cite{Belanger:2012zr,Belanger:2012vp} when $N \geq 2$.

The complex scalar singlet models under $Z_3$ symmetry  have been discussed a lot.
For example, Refs.\cite{Ko:2014nha,Bernal:2015bla} study the SIMPs 
by introducing an extra nonzero vacuume expectation value (VEV) of dark Higgs field with $Z_3$ symmetry
and provide DM candidates in the mass range of ${\cal O}(1-100)$ MeV.
Ref.\cite{Hektor:2019ote} obtains an improved mass bounds
by the study on $Z_3$ singlet dark matter with the refined unitarity bounds and treatment of early kinetic decoupling. These works indicate that two DM mass ranges, $(56.8\sim 58.4)\ {\rm GeV} \leq m_{\rm DM} \leq 62.8\ {\rm GeV}$ and $m_{\rm DM} \geq 122\ {\rm GeV}$, 
are permitted when the semi-annihilation processes play an important role during the freeze-out. In this work, we focus on the study of heavy DM case, say, annihilation of DM particles with masses larger than ${{\cal O}(1)}$ TeV. 

On the other hand, we consider the type-II seesaw mechanism to explain the origin of the tiny neutrino masses. We introduce a $SU(2)$ scalar triplet $\Delta$ with $Z_3$ charge $X_{\Delta}=0$.
This triplet state will obtain a small nonzero VEV after electroweak symmetry breaking (EWSB),
and leads to the Majorana mass origin of neutrinos through Yukawa couplings of leptons and triplets.
A further reason to consider the type-II seesaw is that the introduced  triplet can play an important role in exploring the observed 
excess of cosmic-ray observed in the electron-positron flux mesaured by the AMS-02 \cite{Aguilar:2014mma},
Fermi-LAT \cite{Abdollahi:2017nat} and DAMPE \cite{Ambrosi:2017wek} experiments, 
due to the leptonic decays of such triplet in DM annihilation \cite{Dev:2013hka,Li:2017tmd,Li:2018abw}. Note that the leptophilic dark matter (LDP) mechanism,  
in which a pair of DM particles mainly annihilate into a pair of triplet,
is often adopted to fit the excess of electron-positron spectrum \cite{Dev:2013hka,Li:2017tmd}, where the coupling between DM and Higgs is naturally negligible, leading to negligible semi-annihilation effects. However, the semi-annihilation effects may make sense due to the remarkable characteristics of $Z_3$ symmetry in our model.
We therefore deliberately alleviate the LDP mechanism and reinforce the semi-annihilation effects
to see how they would be constrained accordingly.
As can be seen, the W and Z boson pair from s-channel DM annihilation with subsequent decay 
may lead to inappropriate antiproton spectrum measured by AMS\cite{Aguilar:2002ad} and PAMELA\cite{Adriani:2010rc} in comic-ray,
therefore constrains the $\lambda_{SH}$ coupling between DM and Higgs. It is worth streesing that the boost factor (BF) is also necessary to be considered which may come from large inhomogeneities in the dark matter distribution or due to the so-called Breit Wigner enhancement mechanism in particle physics \cite{Ibe:2008ye, MarchRussell:2008tu,Guo:2009aj}.

The paper is organized as follows. 
In Sec.\ref{sec:model}, we set up the model framework including the gauge, the Yukawa and the scalar sectors.
In Sec.\ref{sec:constraint}, we derive the theoretical constrains, especially the globality of the $Z_3$ symmetry vacuum is studied. 
In Sec.\ref{sec:pheno}, we provide the detailed phenomenological studies and present our numerical results.
Finally, a short summary is given in Sec.\ref{sec:sum}.

\section{The model framework}
\label{sec:model}
We extend the SM by introducing a singlet scalar $S$ (stabilized by a $Z_3$ symmetry), which can be considered as the dark matter candidate.
The scalar triplet $\Delta$ of hypercharge Y=2 is also added to this model  to generate the masses of neutrinos.
Considering that the new $Z_3$ symmetry keeps the scalar potential invariant,
the corresponding transformations are: $H\rightarrow H$, $\Delta\rightarrow \Delta$, $S\rightarrow e^{i 2 \pi/3}S$.
We choose $Z_3$ charge of $S$ equal $X_{S}$=1, and the others with $Z_3$ charges zero.

The total Lagrangian of the model can be written as
\begin{eqnarray}
{\cal{L}}_{\mathrm{tot}}={\cal{L}}_{\mathrm{Kinetic}} + {\cal{L}}_{\mathrm{Yukawa}} - {\cal{V}}(H,\Delta,S) ,
\end{eqnarray}
with the kinetic and Yukawa terms are
\begin{eqnarray}\label{kinetic}
{\cal{L}}_{\mathrm{Kinetic}} &=& (D^{\mu}H)^{\dagger} D_{\mu}H +(D^{\mu}\Delta)^{\dagger} D_{\mu}\Delta + (\partial^\mu{S})^{\dagger}\partial_{\mu}S\ , \\
{\cal{L}}_{\mathrm{Yukawa}} &=& {\cal{L}}^{\mathrm{SM}}_{\mathrm{Yukawa}} - \frac{Y_{ij}}{2} L^{T}_i {\cal{C}} i\sigma_2 \Delta L_j + \mathrm{h.c.}\ .
\end{eqnarray}
Here $Y_{ij}$ represents the Yukawa coupling, $L_{i,j}$ are the $SU(2)_L$ doublet of left-handed leptons,
$i$, $j$ are the generation index and $\cal{C}$ is the charge conjugation operator.
There are no couplings between the scalar singlet $S$ and the SM fermions.
The scalar potential ${\cal{V}}$ will be discussed in detail later.
$H$ and $\Delta$ are labels of the Higgs doublet and scalar triplet respectively which are represented as
\begin{eqnarray}
H&=&\begin{pmatrix}{G^+}\\{\frac{1}{\sqrt{2}}(v_0+h+iG^0)}
\end{pmatrix}\ , \\
\Delta&=&\begin{pmatrix}{\frac{1}{\sqrt{2}}\delta}^{+} & {\delta}^{++}\\\frac{1}{\sqrt{2}}(v_{\Delta}+{\delta}^0+i\eta^{0}) &-\frac{1}{\sqrt{2}}\delta^{+}
\end{pmatrix} \ \ {\mathrm{or}}\ \
\begin{pmatrix}
\delta^{++}\\
\delta^{+}\\
 \frac{1}{\sqrt{2}}(v_\Delta+\delta^0+i \eta^0)
\end{pmatrix}
\end{eqnarray}
with $v_0$ ($v_{\Delta}$) is the VEV of $H$ ($\Delta$).
$G^0$, $G^\pm$ are the Goldstone bosons which are eaten up to give mass to SM gauge bosons.

 The gauge part and Yukawa part  of the model are given in App~.\ref{appA}, and in this section we focus on the scalar part of the model.
The general scalar potential is given by,
\begin{eqnarray} \nonumber
{\cal{V}}(H,\Delta,S) 
 &=& m_h^2 H^\dagger H + \lambda|H|^4 + M^2_{\Delta} \mathrm{Tr}(\Delta^\dagger\Delta) + \mu^2_S S^\dagger S 
  + \lambda_S(S^\dagger S)^2 \\ \nonumber
 &+&[\mu_1 (H^Ti{\sigma}^2\Delta^{\dagger}H)+ \mathrm{h.c.}]
  + \lambda_1(H^\dagger H) \mathrm{Tr}(\Delta^\dagger \Delta) + \lambda_2(\mathrm{Tr}\Delta^\dagger\Delta)^2 \\ \nonumber
 &+& \lambda_3 \mathrm{Tr}(\Delta^\dagger \Delta)^2+\lambda_4 H^\dagger\Delta\Delta^\dagger H + \lambda_{SH} |S|^2|H|^2 \\
 &+& \lambda_{S\Delta}|S|^2 \mathrm{Tr}(\Delta^{\dagger}\Delta)+\frac{\mu_3}{2}(S^3+S^{\dagger 3}),\label{1}
\end{eqnarray}
which can be split into two parts:
\begin{eqnarray}
\mathcal{V}(H,\Delta,S) = \mathcal{V}_{\mathrm{nonDM}}+\mathcal{V}_{\mathrm{DM}}
\end{eqnarray}
with
\begin{eqnarray}
\mathcal{V}_{\mathrm{DM}} =  \mu^2_S S^\dagger S + \lambda_S (S^\dagger S)^2 
+  \lambda_{SH} |S|^2|H|^2+\lambda_{S\Delta}|S|^2 {\mathrm{Tr}}(\Delta^{\dagger}\Delta)+\frac{\mu_3}{2}(S^3+S^{\dagger 3}) \label{DML},
\end{eqnarray}
where the cubic term $(S^3+S^{\dagger3})$ keeps $Z_3$ symmetry invariance 
through $(S^3+S^{\dagger3}) \to ((e^{i 2 \pi /3}S)^3+(e^{-i 2 \pi /3}S^\dagger)^3)$ transformation.
The parameter $\mu_3$ can be regarded as real, owing to its phase can be absorbed into the phase of the singlet $S$,
and not necessary to be negative due to ${\frac{\mu_3}{2}}(S^3+S^{\dagger3}) \to {-\frac{\mu_3}{2}}((-S)^3+(-S)^{\dagger3})$ changing invariance.
The other parameters are also regarded as real for ignoring the CP related issues.
When the EWSB is triggered by the condition of $\mu^2 < 0$,
the SM doublet $H$ obtain a VEV $v_0 \approx 246.22$ GeV and we fix the Higgs mass to be $m_h=125$ GeV.

In the following, we summarize the key formulas we use in our model implementation, 
while recommend Ref.\cite{Arhrib:2011uy} for a more detailed description. 
When the doublet $H$ and the triplet $\Delta$ get the VEVs, we obtain
\begin{eqnarray}
{\cal{V}}(v_0,v_\Delta,0) = \frac{m_h^2}{2} v^2_{0} + \frac{\lambda}{4}v^4_0  + \frac{ M^2_{\Delta}}{2} v^2_{\Delta} + \frac{\lambda_1+\lambda_4}{4}v^2_0 v^2_{\Delta}
+\frac{\lambda_2+\lambda_3}{4} v^4_\Delta - \frac{\mu_1}{\sqrt{2}}v^2_0 v_\Delta .
\end{eqnarray}
By solving the minimal condition of $\partial{\cal{V}}(v_0,v_{\Delta},0)/\partial{v_{\Delta}}=0$ 
and $\partial{\cal{V}}(v_0,v_{\Delta},0)/\partial{v_0}=0$ under the condition $v_{\Delta} \ll v_0$, we can get
\begin{eqnarray}\label{Gv2}
 v_{0}=\sqrt{\frac{-\mu^2}{\lambda}},\ \ v_{\Delta}\approx \frac{\mu_1 v^2_0}{\sqrt{2}(M^2_{\Delta}+\frac{\lambda_1+\lambda_4}{2}v^2_0)}.
%m_h^2 &=& -v^2_0 \lambda - \frac{\lambda_1+\lambda_4}{2} v_\Delta^2 + \sqrt{2} v_\Delta \mu_1 ,\\ 
%m_\Delta^2 &=& -\frac{\lambda_1+\lambda_4}{2} v^2_0 - v^2_\Delta (\lambda_2+\lambda_3) + \frac{\mu_1}{\sqrt{2}} \frac{v^2_0}{v_\Delta} .
\end{eqnarray}

The value of $\mu_1$ is small in the scheme of $\mu_1 \sim v_{\Delta}$ 
so that we can neglect the associated effects in DM annihilation.
For the doubly charged scalar masses, we have 
\begin{eqnarray}
M^2_{\delta^{\pm\pm}} = -v^2_\Delta \lambda_3 - \frac{\lambda_4}{2} v^2_0 + \frac{\mu_1}{\sqrt{2}} \frac{v^2_0}{v^2_\Delta} .
\end{eqnarray}
Here and in the following, without confusion, we use the flavor eigenstate symbol to label its mass eigenstate.
The mass squared matrix for the singly charged field can be diagonalized, with one eigenvalue zero corresponding to the charged Goldstone
boson $G^\pm$ while the other corresponds to the singly charged Higgs boson $\delta^\pm$ and can be given by
\begin{eqnarray}
M^2_{\delta^\pm} = - \frac{v^2_0+2 v^2_\Delta}{4 v_\Delta} ( v_\Delta \lambda_4 - 2 \sqrt{2} \mu_1 ) .
\end{eqnarray}
When the neutral scalar mass matrice is diagonalized, one obtains two massive even-parity physical
states $h$ and $\delta^0$ with the masses given by the eigenvalues
\begin{eqnarray}
M^2_{h} &=& \frac{1}{2}(A + B - \sqrt{(A-B)^2+4 C^2}), \\
M^2_{\delta^0} &=& \frac{1}{2}(A + B + \sqrt{(A-B)^2+4 C^2}).
\end{eqnarray}
with 
\begin{eqnarray}
A &=& 2 v^2_0 \lambda, \\
B &=& 2 v^2_\Delta ( \lambda_2 + \lambda_3) + \frac{\mu_1}{\sqrt{2}}\frac{v^2_0}{v_\Delta}, \\
C &=& v_0 ( v_\Delta (\lambda_1+\lambda_4) - \sqrt{2} \mu_1 ) .
\end{eqnarray}
The pseudoscalar mass matrices leads to one massless Goldstone boson $G^0$ and one massive physical state $\eta^0$
\begin{eqnarray}
M^2_{\eta^0} = \frac{v^2_0 + 4 v^2_\Delta}{\sqrt{2}v_\Delta} \mu_1 .
\end{eqnarray}

From the relation listed above we can write the coupling parameters as function of the masses
\begin{eqnarray}
\mu_1 &=& \frac{\sqrt{2}v_\Delta}{v^2_0+4 v^2_\Delta} M^2_{\eta^0}, \\
\lambda &=& \frac{1}{2v^2_0} ( M^2_{h} \cos^2{\beta} + M^2_{\delta^0} \sin^2{\beta} ), \\
\lambda_4 &=& \frac{4}{v^2_0+4v^2_\Delta} M^2_{\eta^0} - \frac{4}{v^2_0 + 2 v^2_\Delta} M^2_{\delta^{\pm}}, \\
\lambda_3 &=& \frac{1}{v^2_\Delta} \left( \frac{-v^2_0}{v^2_0+4 v^2_\Delta}M^2_{\eta^0} + \frac{2v^2_0}{v^2_0+2 v^2_\Delta} M^2_{\delta^\pm} - m^2_{\delta^{\pm\pm}}  \right), \\
\lambda_2 &=& \frac{1}{v^2_\Delta} \left( \frac{\sin^2\beta M^2_{h} 
+ \cos^2\beta M^2_{\delta^0}}{2} + \frac{1}{2} \frac{v^2_0}{v^2_0+4 v^2_\Delta} M^2_{\eta^0} - \frac{2 v^2_0}{v^2_0+2 v^2_\Delta} M^2_{\delta^\pm} + M^2_{\delta^{\pm\pm}}     \right),\\
\lambda_1 &=& - \frac{2}{v^2_0 + 4 v^2_\Delta} M^2_{\eta^0} + \frac{4}{v^2_0 + 2 v^2_\Delta} M^2_{\delta^{\pm}} + \frac{\sin{2\beta}}{2v_0v_\Delta}( M^2_{h} - M^2_{\delta^0} ).
\end{eqnarray}
with the mixing angle $\beta$ satisfies
\begin{eqnarray}
\sin(2\beta) &=& \frac{4 v_0 \left[ -5 \left( 4 M^2_\Delta + 2 (M^2_{h}+M^2_{\delta^0}) + M^2_h \right) v^2_\Delta \left(v^2_0 + 4 v^2_\Delta\right) + M^2_{\eta} \left(4v^4_0 + 6 v^2_0 v^2_\Delta + 5 v^4_\Delta\right) \right]}
{5 (M^2_{h}-M^2_{\delta^0}) v_\Delta (4 v^2_0 + v^2_\Delta) (v^2_0 + 4 v^2_\Delta)}.~~~~~~~~ 
%\sin^2\beta &=& \frac{ 5(v^2_0+4 v^2_\Delta)( 2(2 M^2_{h}+ M^2_h)v^2_0 - (2 M^2_\Delta + M^2_{\delta^0})v^2_\Delta ) + M^2_A (2 v^4 - 7 v^2_0 v^2_\Delta) }
%{5 (M^2_{h}-M^2_{\delta^0}) (4 v^2_0 + v^2_\Delta) (v^2_0 + 4 v^2_\Delta)} .
\end{eqnarray}
Finally for the dark matter part, we obtain
\begin{eqnarray}\label{Gv3}
M^2_S=\mu^2_S + \frac{\lambda_{SH}}{2} v^2_0 + \frac{\lambda_{S\Delta}}{2} v^2_\Delta.  
\end{eqnarray}
Some other basic relations that may useful are also listed here: 
$g = e/s_W$, $e = \sqrt{4\pi\alpha_{ew}}$,
$s^2_W = \pi\alpha_{ew}/\sqrt{2}/G_f/M^2_W$,
$v^2_0 = \sqrt{ 1/\sqrt{2}G_f-2v^2_\Delta}$,
$M_W = \sqrt{ M^2_Z/2 + \sqrt{ M^4_Z/4 - M^2_Z \pi\alpha/\sqrt{2}/G_f } }$.
We therefore choose our inputs as
\begin{eqnarray}
\alpha_{ew}, M_Z, G_f, M_h
\end{eqnarray}
for the SM part and
\begin{eqnarray}
v_\Delta, M_{\delta^{\pm\pm}}, M_{\delta^{\pm}}, M_{\delta^{0}}, M_{\eta^0}
\end{eqnarray}
for the triplet part and
\begin{eqnarray}
M_S, \lambda_{SH}, \lambda_{S\Delta},\mu_3
\end{eqnarray}
for the dark matter part, respectively. We will also take $v_{\Delta}=1~{\rm eV}$ in our following analysis for simplicity.

\section{Constraints}
\label{sec:constraint}

\subsection{Perturbativity}

To illustrate the theoretical bounds from the perturbativity behavior of the dimensionless scalar quartic couplings, 
we follow the definitions in Refs.\cite{Belanger:2014bga,Lerner:2009xg}.
As to the case of an unrotated basis, the vertices from the potential must be less than $4\pi$ to make sure that 
the tree level contributions are larger than the one-loop level quantum corrections. 
This condition will give the constraints on the couplings $\lambda_i$ in the potential, which are listed here: 
\begin{equation}
\begin{aligned} 
&|6\lambda| \leq 4 \pi, |\lambda_1 + \lambda_4 | \leq 4 \pi, |\lambda_1 |  \leq 4 \pi, |\lambda_1 +\frac{\lambda_4}{2} | \leq 4 \pi, 
|6(\lambda_2 +\lambda_3)|  \leq 4 \pi,\\
&|2 \lambda_2|  \leq 4 \pi,
 |2(2\lambda_2 + \lambda_3)|  \leq 4 \pi,
|\sqrt{2} \lambda_3|  \leq 4 \pi,
|\lambda_4|  \leq 4 \pi,\\
&|\lambda_{S\Delta}|  \leq 4 \pi ,
|\lambda_{SH}|  \leq 4 \pi ,
|4 \lambda_{S}|  \leq 4 \pi .
\end{aligned}
\end{equation}

\subsection{Perturbative unitarity}

The tree-level unitarity from two-body scalar-scalar scattering processes gives another bound on the couplings $\lambda_i$ in the potential.
When the collision energy $\sqrt{s}$ becomes larger, the processes will be dominated by the terms of quartic contact interaction.
Although the trilinear couplings contributed to scattering should be included at finite collision energy \cite{2007Unitarity,Hektor:2019ote},
for simplicity, we only calculate unitarity constrains with the scenarios: $s \rightarrow +\infty$.
Then the s-wave scattering amplitudes lay in the perturbative unitarity limit,
give the constrain of the scalar-scalar scattering S-matrix values: $|\rm{Re}{\cal M}_i| \leq \frac{1}{2}$.
The perturbative unitarity in the type-II seesaw model has been studied with decomposing the matrix $S$
by the mutually unmixed sets of channels with definite charge and CP states\cite{Arhrib:2011uy}.
We extend the way of decomposing by considering the $Z_3$ symmetry and $X_S$=1 singlet $S$ introduced in our model.
The matrix $S$ can be decomposed into seven submatrice blocks structured in terms of electric charges
and $Z_3$ charges in the initial/final states. In Appendix \ref{appA}, we display the initial/final states $E_i$
and the corresponding scattering submatrix ${\cal M}_i$. The corresponding eigenvalues $e^j_i$ of each submatrix are then calculated.
The limit from perturbative unitarity on the potential's couplings $\lambda_i$, i.e., $|\rm{Re}{\cal M}_i| \leq \frac{1}{2}$,
infers $|e_i|\leq 8\pi$.

\subsection{Vacuum stability}

When the scalar field becomes larger in any direction of the field space,
the constraint from vacuum stability is necessary since the scalar potential energy has a finite minimum.
%The constrains from vacuum stability owing to the scalar potential energy having a finite minimum necessarily when the scalar fields become large in any direction of the field space.
In other words, the scalar potential must have a lower bound. 
The quadratic and cubic terms in the scalar potential can be ignored compared with the quartic term in this limit.
These constraints can be achieved by writing the matrix of the quartic interaction on the basis of non-negative field variables
and ensuring that the matrix ${\cal M}$ is copositive \cite{Belanger:2014bga,Kannike:2012pe}.

To parametrize the fields, we can define \cite{Arhrib:2011uy,Belanger:2014bga}:
\begin{equation}
\left\{
\begin{aligned}
            & H^{\dagger}H=r_1^2, \\
            & S=se^{i\phi}, \\
            & Tr(\Delta^{\dagger}\Delta)=r_2^2, \\
            & Tr(\Delta^{\dagger}\Delta)^2/(Tr\Delta^{\dagger}\Delta)^2 \equiv n_1, \qquad n_1 \in[\frac{1}{2},1],\\
            & (H^{\dagger} \Delta \Delta^{\dagger} H)/(H^{\dagger} H Tr\Delta^{\dagger}\Delta) \equiv n_2, \qquad n_2 \in[0,1].\label{Gv}
\end{aligned}
\right.
\end{equation}
The scalar potential about vacuum stability can be written:
\begin{equation}
\begin{aligned}
{\cal V}(H,\Delta,S)_{\rm quartic} &=\lambda|H|^4+\lambda_S(S^\dagger S)^2
+\lambda_1(H^\dagger H)Tr(\Delta^\dagger \Delta)+\lambda_2(Tr\Delta^\dagger\Delta)^2 + \\ 
&\lambda_3 Tr(\Delta^\dagger \Delta)^2+\lambda_4 H^\dagger\Delta\Delta^\dagger H+\lambda_{SH} |S|^2|H|^2+\lambda_{S\Delta}|S|^2 Tr(\Delta^{\dagger}\Delta) \\
%&=(r_1^2,r_2^2,s^2)
%{
%\begin{pmatrix}
%\lambda&\frac{\lambda_1+n_2 \lambda_4}{2}&\frac{\lambda_{SH}}{2}\\
%\frac{\lambda_1+n_2 \lambda_4}{2}&\lambda_2+\lambda_3 n_1&\frac{\lambda_{S\Delta}}{2}\\ 
%\frac{\lambda_{SH}}{2}&\frac{\lambda_{S\Delta}}{2}&\lambda_S
%\end{pmatrix}
%}
%{
%\begin{pmatrix}
%r_1^2\\r_2^2\\s^2
%\end{pmatrix}
%}\\
&=(r_1^2,r_2^2,s^2)
{
\cal M
}
{
\begin{pmatrix}
r_1^2\\r_2^2\\s^2
\end{pmatrix}
},
\end{aligned}
\end{equation}
where
\begin{eqnarray}
{\cal M} =
{
\begin{pmatrix}
\lambda&\frac{\lambda_1+n_2 \lambda_4}{2}&\frac{\lambda_{SH}}{2}\\
\frac{\lambda_1+n_2 \lambda_4}{2}&\lambda_2+\lambda_3 n_1&\frac{\lambda_{S\Delta}}{2}\\ 
\frac{\lambda_{SH}}{2}&\frac{\lambda_{S\Delta}}{2}&\lambda_S
\end{pmatrix}
}
\end{eqnarray}
is a $3\times 3$ symmetric matrix. In Refs.\cite{1983On,1994Nonnegative},
the necessary and sufficient conditions for the matrix $\cal M$ to be copositive had been considered.
Then, the vacuum stability conditions are given below:
\begin{align} \nonumber
 & \lambda \geq 0,\ 
  \lambda_2 + \lambda_3 n_1 \geq 0,\
  \lambda_S \geq 0,\
  \frac{\lambda_1 + n_2 \lambda_4}{2} +\sqrt{\lambda (\lambda_2 + \lambda_3 n_1)} \geq 0, \\ \nonumber
 &  \frac{\lambda_{SH}}{2} + \sqrt{\lambda \lambda_{S}} \geq 0,\
  \frac{\lambda_{S \Delta}}{2} +  \sqrt{(\lambda_2 +\lambda_3 n_1) \lambda_S} \geq 0, \\ \nonumber
 &\sqrt{\lambda (\lambda_2 + \lambda_3 n_1) \lambda_S}
   +\frac{\lambda_1 + n_2\lambda_4}{2} \sqrt{\lambda_{S}}
   +\frac{\lambda_{SH}}{2} \sqrt{\lambda_2 + \lambda_3 n_1}
   +\frac{\lambda_{S\Delta}}{2} \sqrt{\lambda}
   +\\&\sqrt{2 (\frac{\lambda_1 + n_2 \lambda_4}{2} +
   \sqrt{\lambda (\lambda_2 + \lambda_3 n_1)})( \frac{\lambda_{SH}}{2} 
   + \sqrt{\lambda \lambda_{S}})( \frac{\lambda_{S \Delta}}{2} 
   +  \sqrt{(\lambda_2 +\lambda_3 n_1) \lambda_S})} \geq 0.
\end{align}
Here, $n_1 \in[\frac{1}{2},1]$ and $n_2 \in[0,1]$.

\subsection{Globality of the $Z_3$-symmetric vacuum}

Since we choose the complex singlet scalar $S$ as the DM candidate, the $Z_3$ symmetry should remain unbroken.
The vacuum stability condition leads to the existence of a finite global minimum in the scalar potential.
To ensure that the SM $(\bcancel{EW}, Z_3)$ vacuum is selected as the global minimum vacuum,
we study the stationary points at the extrema of the  scalar potential.
Following the parametrizing of the fields in Eq.(\ref{Gv}), the stationary points can be obtained by
taking the derivative of the potential ${\cal V}(r_1,r_2,s,\phi)$
with respect to $r_1$, $r_2$, $s$ and $\phi$ respectively and solving the equations of
\begin{eqnarray}\label{stapoi} \nonumber
0 &=& r_1(2\lambda r^2_1+\lambda_1 r^2_2+\lambda_4 n_2 r^2_2 +\lambda_{SH} s^2 +m^2_h),\\ \nonumber
0 &=& r_2(\lambda_1 r^2_1 +2 \lambda_2 r^2_2 + 2 \lambda_3 n_1 r^2_2+M^2_{\Delta}+\lambda_4 n_2 r^2_1 +\lambda_{S\Delta}s^2),\\ \nonumber
0 &=& s(4\lambda_S s^2+2\mu^2_S+2\lambda_{SH}r^2_1+2\lambda_{S\Delta}r^2_2+3\mu_3 s \cos(3\phi)), \\
0 &=& s^3\mu_3 \sin(3\phi).
\end{eqnarray}
Here, we have ignored $\mu_1$ ($\mu_1 \sim \mu_{\Delta}$). We use Eq.(\ref{Gv2}-\ref{Gv3}) to simplify the form of solution of Eq.(\ref{stapoi}).
The chosen scheme of triplet scalar mass degeneracy leads to $\lambda_3\simeq 0$ and $\lambda_4\simeq 0$.
And since $v_{\Delta}\ll v_{0}$ so that $\lambda_1 \approx \lambda_2 \approx \frac{2M^2_h}{v_0^2}$.
Due to the chosen condition of $\lambda_{S\Delta} > 0$, and $M_{\Delta}^2 > 0$, we have $r_2=0$.
We also set $\mu_3 \geq 0$ and $\cos 3\phi = -1$  to obtain a local minima of potential with $s \ne 0$ \cite{Belanger:2012zr}.
Finally, there are only four vacuums left that should be considered, we give the discussion below:
\begin{description}
\item 1. ($r_1=0$, $s=0$) vaccum: electroweak (EW) and $Z_3$ symmetries keep unbroken, $v_h=v_{\Delta}=v_s=0$,
\begin{equation}
{\cal V}_{(EW, Z_3)} = {\cal V}(r_1, r_2, s, \phi)\bigg|_{r_1=\frac{v_{h}}{\sqrt{2}}, r_2=\frac{v_{\Delta}}{\sqrt{2}}, s=v_s, \phi=\frac{1}{3}arcos(-1)} = 0.
\end{equation}
\item 2. ($r_1 \neq 0$, $s=0$) vaccum: EW symmetry broken and $Z_3$ symmetry is retained with $v_h^2 \approx v_0^2$, $v_{\Delta} \approx 0$, $v_s=0$,
\begin{equation}
{\cal V}_{(\bcancel{EW}, Z_3)} = {\cal V}(r_1, r_2, s, \phi)
\bigg|_{r_1=\frac{v_{h}}{\sqrt{2}}, r_2=\frac{v_{\Delta}}{\sqrt{2}}, s=v_s, \phi=\frac{1}{3}arcos(-1)} \approx -\frac{(M_h v_0)^2}{8}.
\end{equation}
\item 3. ($r_1=0$, $s \neq 0$) vaccum: $Z_3$ symmetry broken and EW symmetry is retained with the condition
\begin{equation}
v_h=v_{\Delta}=0, v_s=Root\left[\frac{\partial{{\cal V}(r_1, r_2, s, \phi)}}{\partial{s}}\bigg|_{r_1=0, r_2=0,\phi=\frac{1}{3}arcos(-1)}=0\right].
\end{equation}
Here,the symbol Root denotes the solution of  partial differential equation. Thus,
\begin{equation}\label{V3}
{\cal V}_{(EW, \bcancel{Z_3})} = {\cal V}(r_1, r_2, s, \phi)\bigg|_{r_1=\frac{v_{h}}{\sqrt{2}}, r_2=\frac{v_{\Delta}}{\sqrt{2}}, s=v_s, \phi=\frac{1}{3}arcos(-1)}.
\end{equation}
\item 4. ($r_1 \neq 0$, $s \neq 0$) vaccum: Breaking both EW and $Z_3$ symmetries  with the condition
\begin{equation}
v_{\Delta}\approx 0,\
\begin{pmatrix}
v_h\\
v_s
\end{pmatrix}
=Root\left[
\begin{pmatrix}
\frac{\partial{{\cal V}(r_1, r_2, s, \phi)}}{\partial{h}}\bigg|_{r_2=0,\phi=\frac{1}{3}arcos(-1)}=0 \\
\frac{\partial{{\cal V}(r_1, r_2, s, \phi)}}{\partial{h=s}}\bigg|_{r_2=0,\phi=\frac{1}{3}arcos(-1)}=0
\end{pmatrix}
\right].
\end{equation}
Thus,
\begin{equation}\label{V4}
{\cal V}_{(\bcancel{EW}, \bcancel{Z_3})} = {\cal V}(r_1, r_2, s, \phi)
\bigg|_{r_1=\frac{v_{h}}{\sqrt{2}}, r_2=\frac{v_{\Delta}}{\sqrt{2}}, s=v_s, \phi=\frac{1}{3}arcos(-1)}.
\end{equation}
\end{description}

\begin{figure}[htbp]
\centering
\subfigure[]{\includegraphics[height=4.9cm,width=4.9cm]{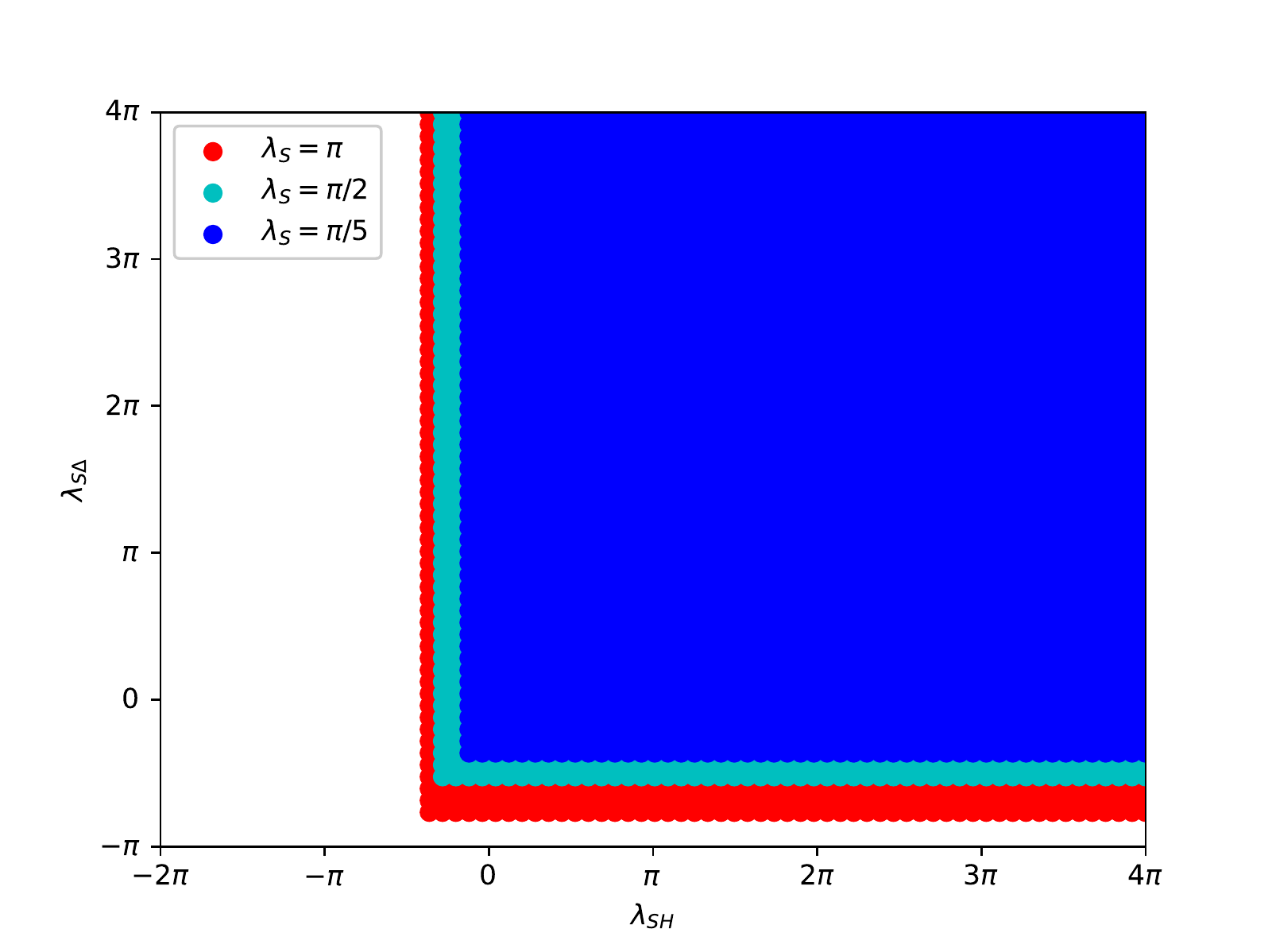}}
\subfigure[]{\includegraphics[height=4.9cm,width=4.9cm]{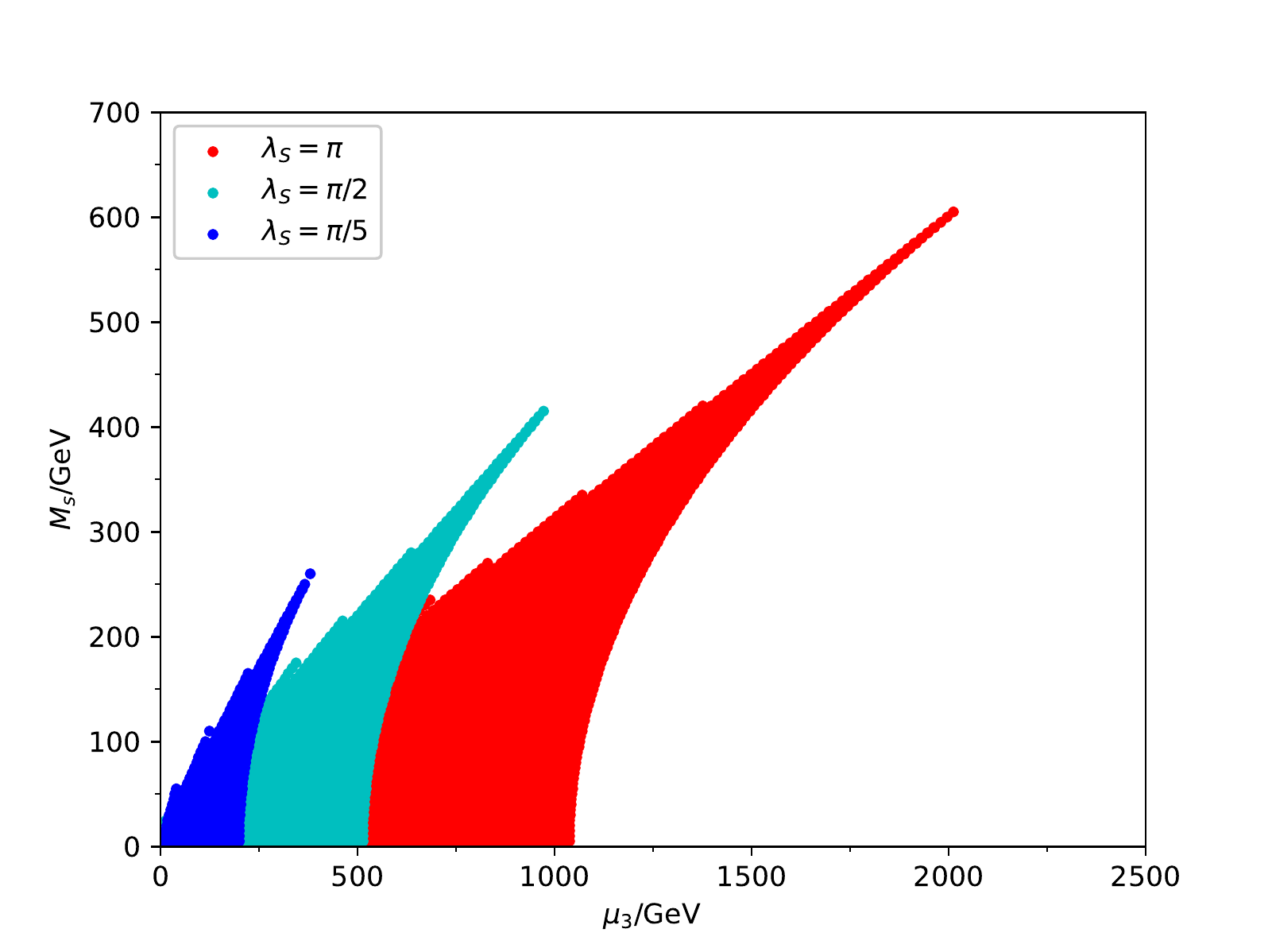}}
\subfigure[]{\includegraphics[height=4.9cm,width=4.9cm]{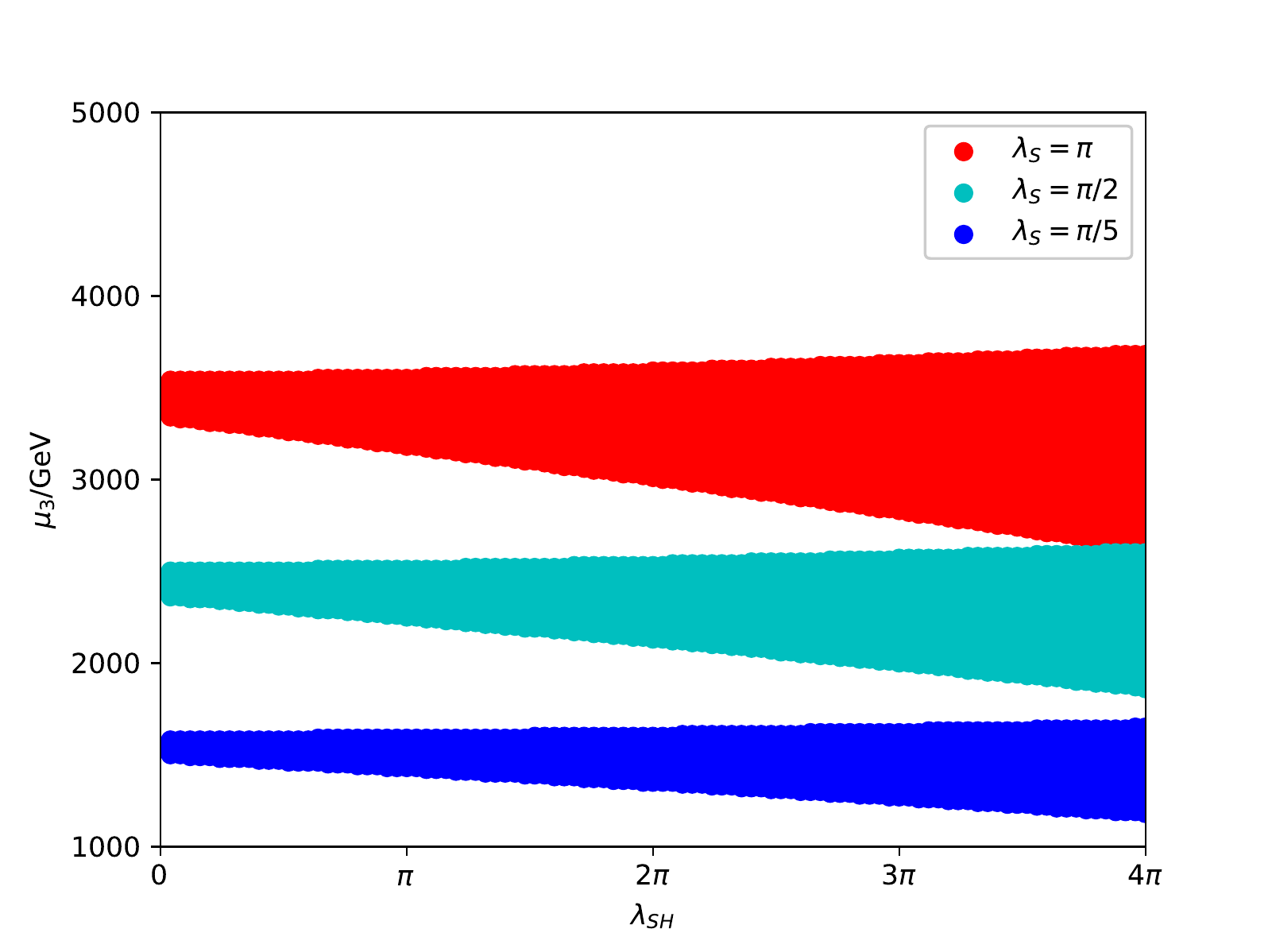}}
\caption{
(a) The possible area in $\lambda_{SH}$ vs $\lambda_{S\Delta}$ plane from perturbativity, perturbative unitary and vacuum stability
by taking $\lambda_3 \approx \lambda_4 \approx 0$ and $\lambda_1 \approx \lambda_2\approx \frac{2M^2_h}{v^2_0}$ at $\lambda_S=$ $\pi$, $\pi/2$, $\pi/5$.
(b) The possible area by requiring that ${\cal V}_{\bcancel{EW}, Z_3}$ vacuum is the expected SM one.
(c) The possible area in $\lambda_{SH}-\mu_3$ plane with $M_S=1$ TeV.}
\label{glob}
\end{figure}

To ensure the condition that the vacuum value of ${\cal V}_{\bcancel{EW},Z_3}$ is below the others,
we need to estimate the values of Eq.(\ref{V3}) and Eq.(\ref{V4}) at given points $(M_S,\mu_3,\lambda_{S},\lambda_{SH})$.
We make a numerical scan in Figure \ref{glob}.
Firstly, in Figure \ref{glob}(a), we present the allowed area of $\lambda_{SH}$ and $\lambda_{S\Delta}$
from perturbativity, perturbative unitary and vacuum stability constraints,
by taking $\lambda_3\simeq 0$, $\lambda_4\simeq 0$ and $\lambda_1=\lambda_2\approx \frac{2m^2_h}{v^2_0}$.
The red, green and blue areas correspond to $\lambda_S$ equal $\pi$, $\pi/2$ and $\pi/5$, respectively.
It is clear that the constraint of $\lambda_{SH}$ and $\lambda_{S\Delta}$ from these conditions are weak.
Moreover, the allowed area is reduced as the value of $\lambda_S$ becomes small.
Then, we take the region of $\lambda_{SH}, \lambda_{S\Delta} \in [0,4\pi]$ to scan over.
Then, in Figure \ref{glob}(b) we present the scan result by requiring that ${\cal V}_{\bcancel{EW}, Z_3}$ vacuum is the expected SM one.
The allowed region is restricted within the $\mu_3-M_S$ plane.
One comment is that, as $M_S$ becomes larger, the solution of ${\cal V}_{\bcancel{EW}, \bcancel{Z_3}}$ vacuum doesn't exist,
due to the violation of at least one of the condition $v_s > 0$, $v^2_h > 0$, $D_{\bcancel{EW}, \bcancel{Z_3}} \geq 0$ ,
where $D_{\bcancel{EW}, \bcancel{Z_3}}$ is the discriminant of the quadratic form defined in Ref.\cite{Belanger:2012zr}.
Only values of ${\cal V}_{\bcancel{EW}, Z_3}$ and ${\cal V}_{EW,\bcancel{Z_3}}$ need to be compared with each other, resulting in a corresponding limit on $\mu_3$.
This can be further seen clearly in Figure \ref{glob}(c) where we fix $M_S=1$ TeV and vary $\mu_3$.
The upper bounds express the maximum value of $\mu_3$ meet the condition of ${\cal V}_{\bcancel{EW}, Z_3} < {\cal V}_{EW,\bcancel{Z_3}}$,
and the lower boundary indicates the corresponding minimum value of $\mu_3$.
Thus, the requirement that the SM vacuum is the the global minimum gives a limit on the maximum value of $\mu_3$.
Moreover, to obtain a larger $\mu_3$, a larger $\lambda_S$ is demanded.
\section{Phenomenology}
\label{sec:pheno}

\subsection{Relic density and direct detection constraint}

\begin{figure}[htbp]
  \centering
  \subfigure[]{\includegraphics[width=0.40\textwidth,height=3.0cm]{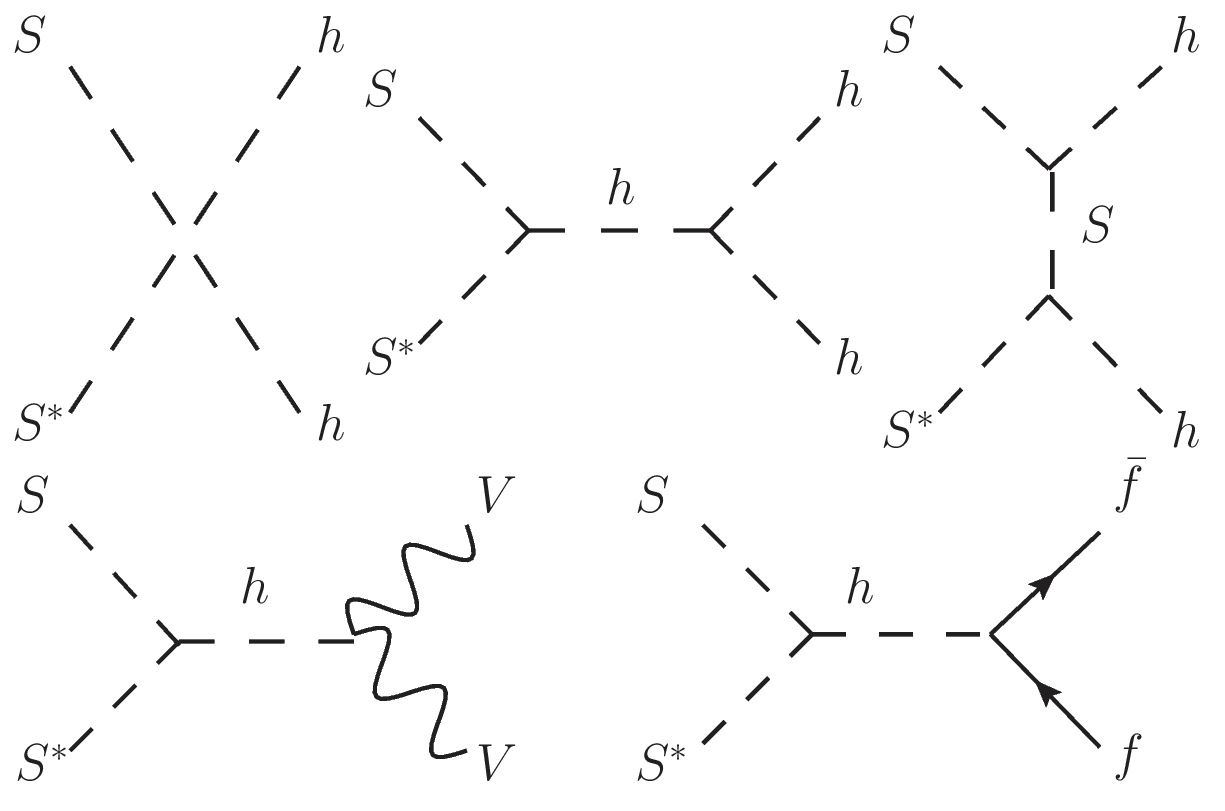}}
  \hspace{8mm}
  \subfigure[]{\includegraphics[width=0.20\textwidth,height=3.0cm]{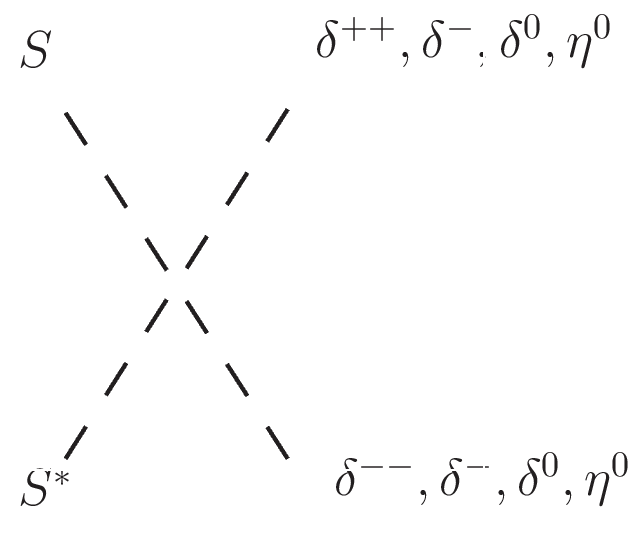}}
  \hspace{3mm}
  \subfigure[]{\includegraphics[width=0.20\textwidth,height=3.0cm]{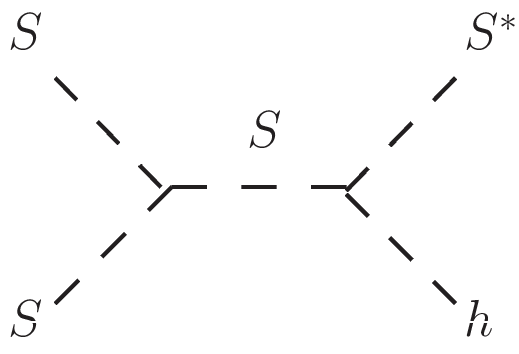}}
  \caption{DM scattering Feynman diagrams of annihilating to the SM particles (a), triplet particles (b),
  and semi-annihilating to a dark matter and a Higgs particle (c).}
\label{feynp}
\end{figure}

We are now considering the contribution of DM thermal cross sections to the relic abundance.
There are mainly three ways of DM scatterings:
(a) annihilation to the SM particles;
(b) annihilation to the triplet particles;
(c) semi-annihilation to a dark matter and a Higgs particle.
See Figure.\ref{feynp} for the illustrated channels in details.

The number density of the DM particles $n$ satisfies the Boltzmann equation:
\begin{eqnarray} \nonumber
\frac{dn}{dt} + 3Hn &=& - \langle \sigma v \rangle^{SS^{*}\to XX}(n^2-{\bar{n}}^2)
                        - \langle \sigma v \rangle^{SS^{*}\to \Delta\Delta}(n^2-{\bar{n}}^2) \\
		        &-& \frac{1}{2} \langle \sigma v \rangle^{SS\to S^{*}h}(n^2-n\bar{n}) \ \ \ \
\end{eqnarray}
where $ \bar{n}$ is the number density in thermal equilibrium, $H$ is the Hubble expansion rate of the Universe,
$X$ denotes SM particles, $\Delta$ are the triplet particles ($\delta^{++(--)}$, $\delta^{+(-)}$, $\delta^{0}$, $\eta^{0}$),
and $\langle \sigma v \rangle$ is the thermally averaged annihilation cross section,
in terms of the partial wave expansion $\langle \sigma v \rangle \simeq a + b v^2$.
To distinguish different contributions of these three scattering channels on thermal cross section,
we define the scattering cross section fraction $N_i(i=1,2,3)$ as:
\begin{eqnarray}
N_1 &=& \frac{\frac{1}{2}\langle v\sigma \rangle^{SS\to S^{*}h}}{ \langle v\sigma \rangle^{SS^{*}\to \Delta\Delta}
       + \langle v\sigma \rangle^{SS^{*}\to XX} + \frac{1}{2}\langle v\sigma \rangle^{SS\to S^{*}h  }} \times 100\% ,  \\
N_2 &=& \frac{ \langle v\sigma \rangle^{SS\to \Delta\Delta}}{ \langle v\sigma \rangle^{SS^{*}\to \Delta\Delta}
       + \langle v\sigma \rangle^{SS^{*}\to XX} + \frac{1}{2}\langle v\sigma \rangle^{SS\to S^{*}h }}\times 100\% , \\
N_3 &=& (1-N_1-N_2).
\end{eqnarray}
Here $N_3$ denotes the fraction of DM pairs annihilating to SM particles.
To calculate the DM relic density and the fraction $N_i$ we use the micrOMGEAs5.0.6 package \citep{Belanger:2018ccd},
in which the model has been implemented through the FeynRules package \cite{Alloul:2013bka}.
%{\color{red} The remaining numerical analysis has been performed by ....}.
To constrain the parameter space, we require the relic density to fit
the $2\sigma$ C.L. range of Planck result \cite{Ade:2015xua}: $\Omega_{DM}h^2=0.1199\pm0.0027$.
%{\color{red}For enhancing the channels a pair of DM annihilates, $M_S>M_{\Delta}$ should be %selected.   ??????before, you always use $m_{\rm DM}$???}
The direct detection constraint is obtained from the spin-independent (SI) elastic scattering measurements.
which can be  given by  \cite{Li:2017tmd}
\begin{equation}
\sigma_{SI}=\frac{\lambda^2_{SH}}{\pi M^4_H}\frac{M^2_N}{(M_N+M_S)^2}\times 0.0706M^2_N,
\end{equation}
where the nucleon mass $M_N\simeq 0.939$ GeV.
The  XENON1T experiment \cite{XENON:2018voc} has given the upper limit of the SI elastic scattering cross section
at a given DM mass $M_S$, which will lead to a maximum value of $\lambda_{SH}$.
As $N_1$ is proportional to $\mu^2_3\lambda_{SH}^2/M^6_S$,
in order to enhance the semi-annihilate contribution characterized by the $Z_3$ singlet scalar $S$,
we choose $\lambda_S=\pi$ so that the parameter $\mu_3$ is as larger as possible,
while the condition that the $(\bcancel{EW}, Z_3)$ vacuum being the global minimum is still guaranteed,
see in Figure \ref{glob}(c) that has been discussed.

\begin{figure}[htbp]
\centering
  \includegraphics[height=6.0cm,width=7cm]{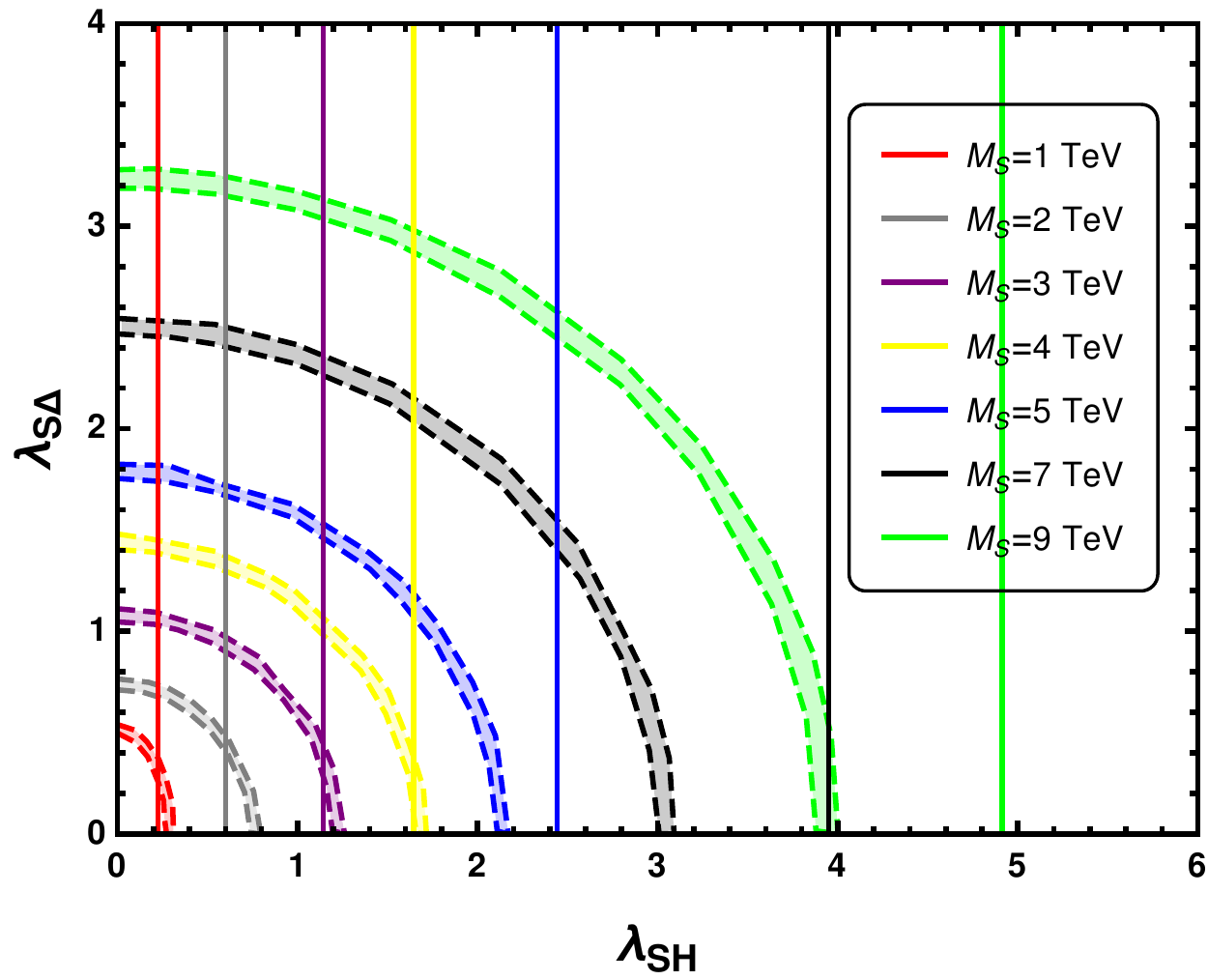}
%  \subfigure[]{\includegraphics[height=6.0cm,width=7cm]{ms3.eps}}
\caption{
The allowed parameter space in $\lambda_{SH}$ vs $\lambda_{S\Delta}$ plane adoptting
the requirement that the relic density should fit the Planck measurement in the 2$\sigma$ C.L. range.
%$M_{\Delta}=0.9$ TeV with changed DM mass $M_S$.
The vertical line is the upper value of $\lambda_{SH}$ at a given $M_S$ based on XENON1T measurements.}
\label{fitOmega}
\end{figure}

%\begin{table}[htpb]
%\centering
%    \begin{tabular}{c|ccccc}
%    \hline
%    \hline
%    $M_S$[TeV] & $\lambda_{SH}$ & $\lambda_{S\Delta}$ & $\mu_3$[GeV] & $N_1$ & $N_2$ \\
%    \hline
%    1 & 0.19(PX.L.) & 0.39 & 3544 & 24\% & 57\% \\
%    \hline
%    2 & 0.54(PX.L.) & 0.52 & 7089 & 12\% & 48\% \\
%    \hline
%   3 & 0.99(PX.L.) & 0.62 & 10637 & 8\% & 32\% \\
%    \hline
%    4 & 1.52(PX.L.) & 0.63 & 14179 & 6\% & 18\% \\
%    \hline
%    5 & 2.13(PX.L.) & 0$\sim$0.34 & 17724 & 5\% & 0\% \\
%    \hline
%   7 & 3.07 & 0$\sim$0.20 & 24814 & 2\% & 0\% \\
%    \hline
%    9 & 3.98 & 0$\sim$0.24 & 31904 & 2\% & 0\% \\
%   \hline
%    \hline
%    \end{tabular}
%\caption{Benchmarks with the maximum $\lambda_{SH}$ in Fig.(\ref{fitOmega}). The symbol PX.L. denotes the PandaX-II limit.}
%\label{benmks}
%\end{table}

A detailed scan is shown in Figure \ref{fitOmega}. As can be seen,
we fix $M_\Delta=0.9$ TeV while vary benchmark values of $M_S$ that obey $M_S > M_{\Delta}$.
Meanwhile, adopt the requirement that the relic density should fit the Planck measurement in the 2$\sigma$ C.L. range,
the parameter space of $\lambda_{SH}$ vs $\lambda_{S\Delta}$ is constrained within a quarter circle ring.
Different ring relates to different choice of $M_S$.
The vertical lines correspond to the XENON1T measurements at a fixed $M_{\Delta}$.
Clearly, as the DM mass $M_S$ becomes larger, the upper boundary of $\lambda_{SH}$ becomes larger correspondingly.
The crossover point (or precisely, segment) of the same coloured ring and line, relates to a boundary point.
The region on the left side of point fits both Planck as well as XENON1T measurements,
while the right side does not fit that of XENON1T.
%As the DM mass $M_S$ become larger, the constrants on $\lambda_{SH}$ from Planck is tighter than that from PandaX-II.
With the increase of DM mass, the restriction of Planck experiment on $\lambda_{SH}$ is more important than that of XENON1T.
Notice when $M_S\geq 4$ TeV, the constaint of XENON1T experiment will make little difference,  and only Planck results make sense.

%More information at the crossover points is shown in Table \ref{benmks}.
%The second column is the upper boundary of $\lambda_{SH}$ that fits both experiments.
%{\color{red}P.L. means ...}. 
%The upper values of $\mu_3$ are also shown in the forth column, 
%where the limits are set by the condition of ${\cal V}_{\bcancel{EW}, Z_3} < {\cal V}_{EW,\bcancel{Z_3}}$.
%The scattering cross section fractions are displayed in the last two columns.
%Though $N_1$ is proportional to $\mu^2_3\lambda^2_{SH}$, it is inversely proportional to the sixth power of $M_S$, 
%so that the maximum value of $N_1$ decreases significantly with the increase of $M_S$, even $\mu_3$ increases.

\subsection{Antiproton spectrum and electron-positron flux}

The DM particles may annihilate to W and Z boson pairs through the s-channel Higgs exchange.
The subsequent decays of the bosons to antiprotons will be responsible for the interpretation of
the cosmic-ray antiprotons spectrum that has been measured by AMS \cite{Aguilar:2002ad}, PAMELA \cite{Adriani:2010rc} collaborations.
On the other hand, the triplets produced from DM (co)annihilation may decay to leptons ($\Delta \rightarrow \ell^i\ell^j$) through Yukawa interactions,
and is therefore possible to account for the excess of electron-positron flux in cosmic-ray exhibited
in the AMS-02, Fermi-LAT, DAMPE experiments \cite{Aguilar:2014mma, Abdollahi:2017nat, Ambrosi:2017wek}.
Notice the excess of position-electron flux in cosmic-ray may also be explained by astrophysical evidence, for example, an isolated young pulsar \cite{Yuan:2017ysv}.
Here in our paper, we focus on the DM interpretation although it is possible that the cosmic ray fluxes are not due to DM but due to the mundane astrophysics. Interpretation of the  cosmic ray excesses by the dark matter has been discussed a lot, and related works can be found in Refs.~\cite{Li:2018abw,Okada:2017pgr,Randall:2019zol,YaserAyazi:2019psw,Kachelriess:2019oqu,Feng:2019rgm,Cappiello:2018hsu,Yuan:2018rys,Chen:2017tva,Liu:2017rgs,Jin:2017qcv}
We choose the DM mass at the 3 TeV scale, which is large enough to produce the lepton with the energy of the order of ${\cal O}(1)$ TeV.
Following our above discussion, we choose $M_{\Delta}=0.9$ TeV and $\mu_3 \approx $ 10 TeV to enhance the semi-annihilation as so as possible.
We focus on the parameter space where our model can give a natural DM explanation of both cosmic ray measurements as well as fitting relic density measurement simultaneously,
which also includes semi-annihilation effects.

To calculate the antiproton flux and electron-positron flux,
we use the following parametrization functions \cite{Bringmann:2006im, Baltz:1998xv, Baltz:2001ir}
\begin{eqnarray}
\label{fanti}
&&\log_{10}\Phi^{bkg}_{\bar{p}}=-1.64+0.07x-x^2-0.02x^3+0.028x^4, \\
\label{eflux1}
&&\Phi^{prim}_{e^-}(E)=\frac{0.16E^{-1.1}}{1+11E^{0.9}+3.2E^{2.15}}[GeV^{-1}cm^{-2}s^{-1}sr^{-1}], \\
&&\Phi^{sec}_{e^{-}}(E)=\frac{0.70E^{0.7}}{1+110E^{1.5}+600E^{2.9}+580E^{4.2}}[GeV^{-1}cm^{-2}s^{-1}sr^{-1}], \\
\label{eflux2}
&&\Phi^{sec}_{e^{+}}(E)=\frac{4.5E^{0.7}}{1+650E^{2.3}+1500E^{4.2}}[GeV^{-1}cm^{-2}s^{-1}sr^{-1}],
\end{eqnarray}
with $x=\log_{10}T/GeV$ in Eq.(\ref{fanti}), the label $\Phi^{bkg}_{\bar{p}}$ denotes the cosmic-ray antiproton background.
The label $\Phi^{prim(sec)}$ in Eq.(\ref{eflux1})-Eq.(\ref{eflux2}) means the primary (secondary) cosmic of electron or positron background.
The formula is appropriate for the energy range 10-1000 GeV \cite{Baltz:1998xv}.
The primary and secondary electron backgrounds are originated from supernova remnants and cosmic ray spallation in the interstellar medium, respectively.
The secondary positrons background comes from primary protons colliding with other nuclei in the interstellar medium.
With the value of BF mentioned above, the total antiproton and positron plus electron flux are given by:
\begin{eqnarray}
&&\Phi_{\bar{p}}=\Phi^{bkg}_{\bar{p}}+BF\times \Phi^{DM}_{\bar{p}}. \\
&&\Phi_{e^+}+\Phi_{e^-}=k(\Phi^{(prim)}_{e^{-}}+\Phi^{(sec)}_{e^{-}}+\Phi^{(sec)}_{e^{+}})+BF \times (\Phi^{DM}_{e^{-}}+\Phi^{DM}_{e^{+}}),
\end{eqnarray}
where $k$ is the normalization parameter and we fix it as 0.9, $\Phi^{DM}$ is the corresponding flux from DM pair annihilation.
The background fluxes can in principle be estimated.
To calculate $\Phi^{DM}$ we use micrOMEGAs, in which the density distribution of DM in the galactic halo is taken from Navarro-Frenk-White (NFW) density,
and the effects of galactic charged particles propagation and solar modulation are considered.

\begin{figure}[htbp]
\centering
\subfigure[]{\includegraphics[height=4.5cm,width=4.9cm]{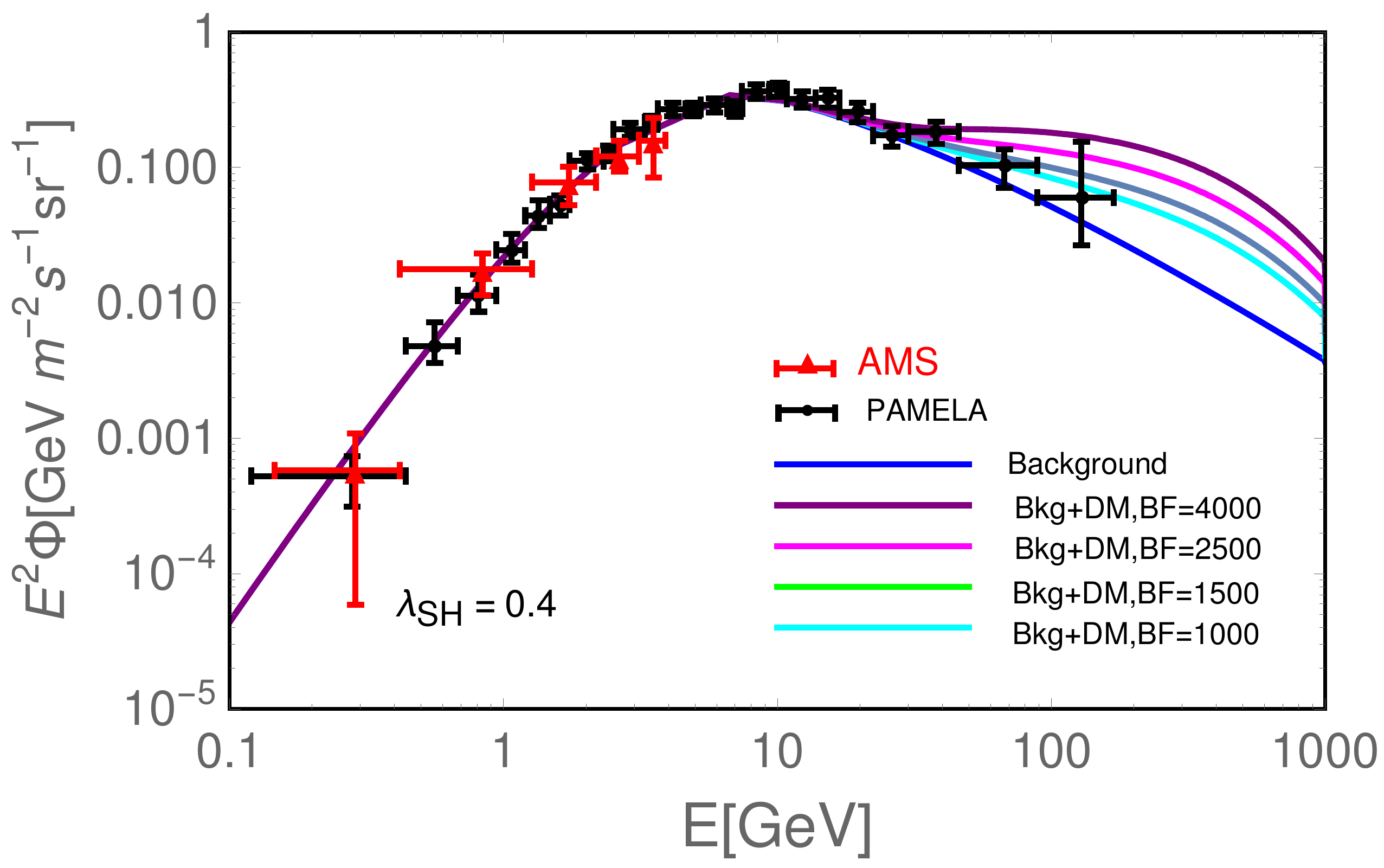}}
\subfigure[]{\includegraphics[height=4.5cm,width=4.9cm]{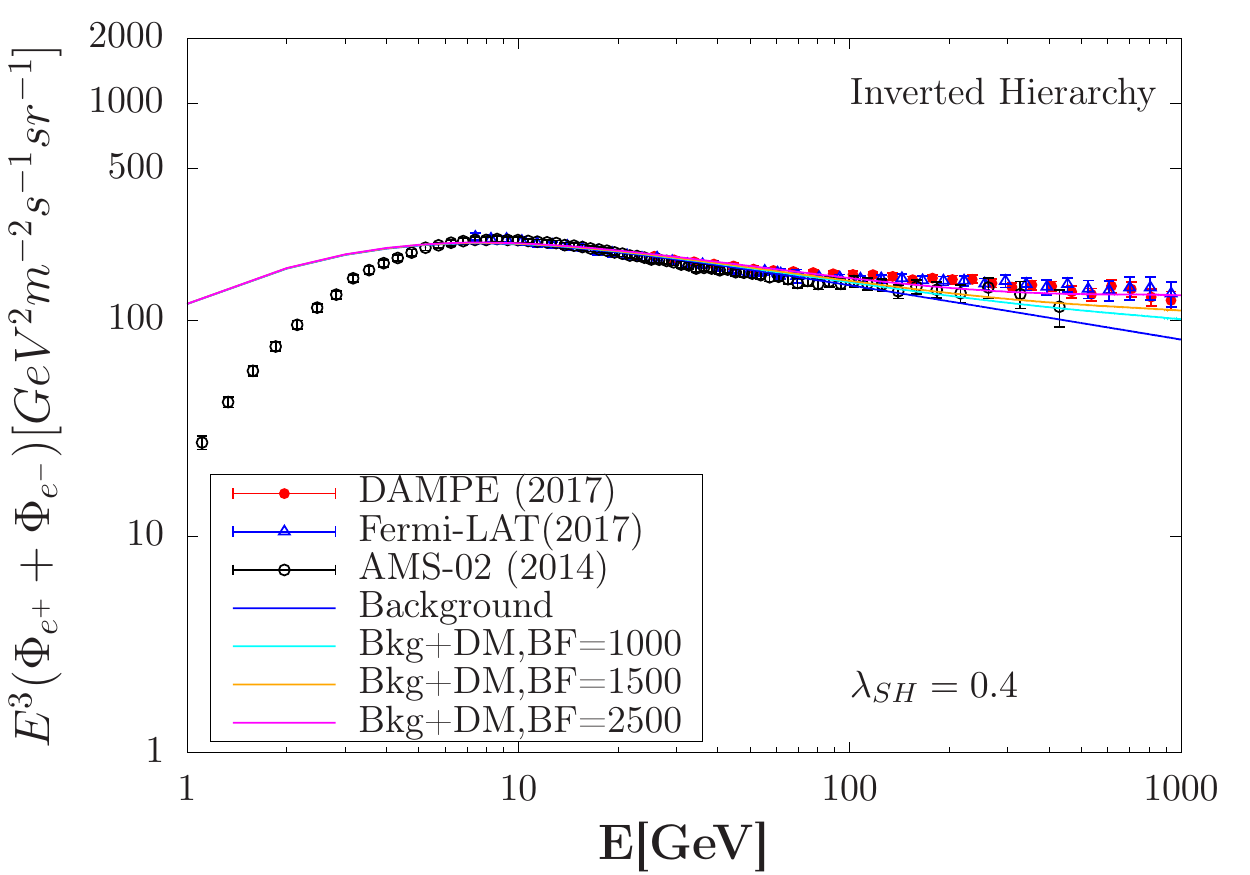}}
\subfigure[]{\includegraphics[height=4.5cm,width=4.9cm]{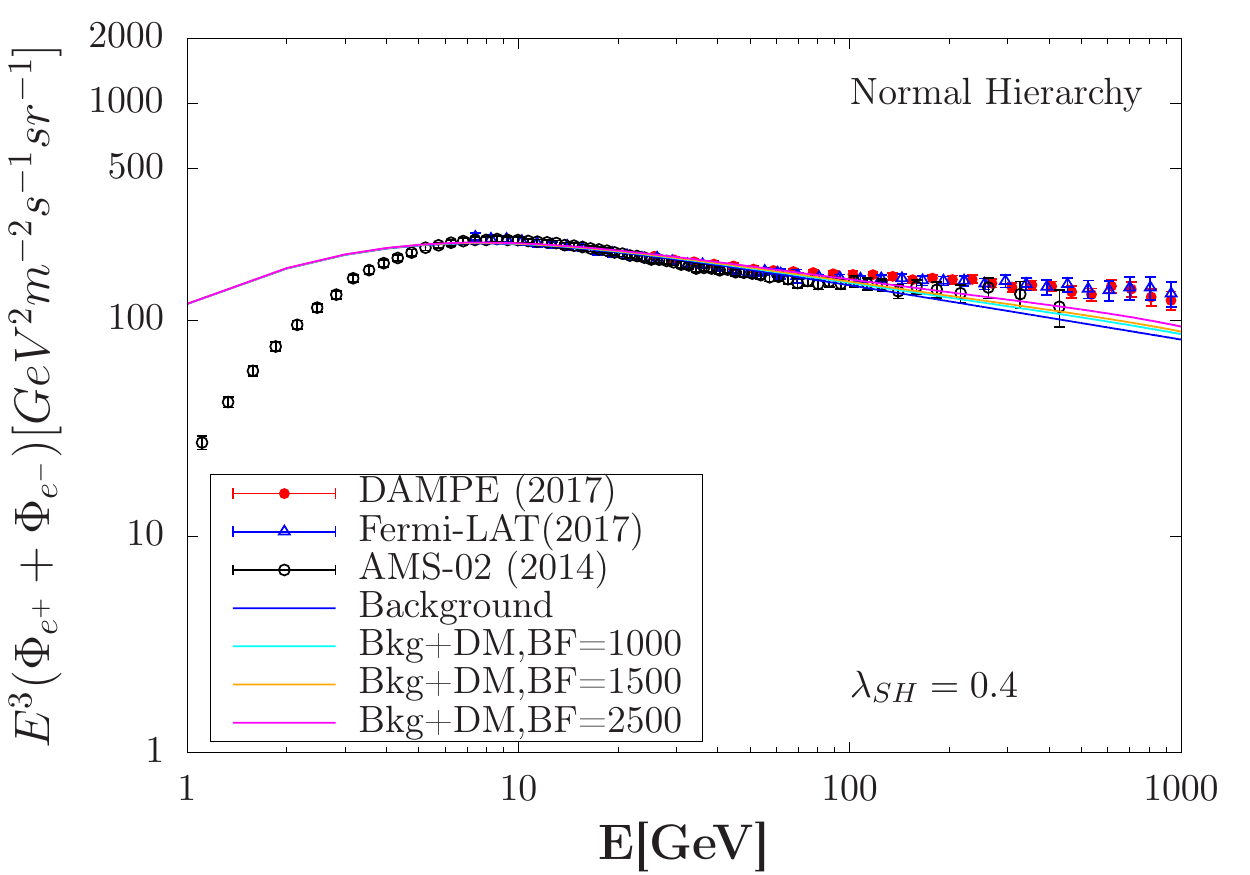}}
\caption{ 
Background and background + DM (Bkg+DM) of cosmic-ray antiproton flux in (a), 
and electron-positron flux in (b) and (c) with inverted hierarchy (IH) and normal hierarchy (NH) scenario, respectively.
%at $\lambda_{SH}$=0.4 and varied BF, 
The data points in (a) labels the PAMELA\cite{Adriani:2010rc} and AMS\cite{Aguilar:2002ad} measurements.
The data points in (b) and (c) stand for the AMS-02\cite{Aguilar:2014mma}, Fermi-LAT\cite{Abdollahi:2017nat} and DAMPE\cite{Ambrosi:2017wek} measurements.}
\label{result_0.4}
\end{figure}

Considering the difference in the electron-positron spectrum between DAMPE and Fermi-LAT measurements when $E>1$ TeV, 
we focus on the calculation of the flux when the value of $E$ is located in the range of $0 \sim 1000$ GeV. 
The results of cosmic-ray antiproton as well as electron-positron fluxes are displayed in Figure \ref{result_0.4} 
at $\lambda_{SH}=0.4$ with different values of BF taken as free parameter.
Figure \ref{result_0.4}(a) displays the background, background + DM (Bkg+DM) for $\Phi_{\bar{p}}$. 
Figure \ref{result_0.4}(b) and (c) show the background, background + DM (Bkg+DM) for $\Phi_{e^+}+\Phi_{e^-}$ in both IH and NH scenarios, respectively.
As can be seen in Figure \ref{result_0.4}(a), 
at a given value of $\lambda_{SH}=0.4$, the PAMELA \cite{Adriani:2010rc} results impose strong restrictions on BF, where the maximum value is about 2500.
Then in Figure \ref{result_0.4}(b) and (c), we keep the same value of $\lambda_{SH}$, vary BF up to the maximum of 2500, 
and compare the positron-electron fluxes with DAMPE, Fermi-LAT and AMS-02 experiments. 
We find that, in the IH scenario, flux with small BF meets AMS and PAMELA results well while the DAMPE and Fermi-LAT experiments favor large values of BF.
Select the appropriate BF value, we can fit all the experiments with our chosen parameters.
The case is different for the NH scenario where the results only meet the AMS-02 measurements.
There displays a suppression in the high $E$ region.
If we believe more in the results of DAMPE and Fermi-LAT experiments, the IH scenario is disfavored.
The different behavior between these two scenarios is understood that, in our case, 
DM particles mainly annihilate into electron final states in the IH scenario rather than tau final states in the NH one.
This can be seen directly from the final state decay fraction in Table \ref{Br},
which is $e:\mu:\tau \approx 1:0.2:0.3$ for the IH scenario and $e:\mu:\tau \approx 1:0.5:1.4$ for that of NH.

\begin{figure}[htbp]
\centering
\subfigure[]{\includegraphics[height=4.5cm,width=4.9cm]{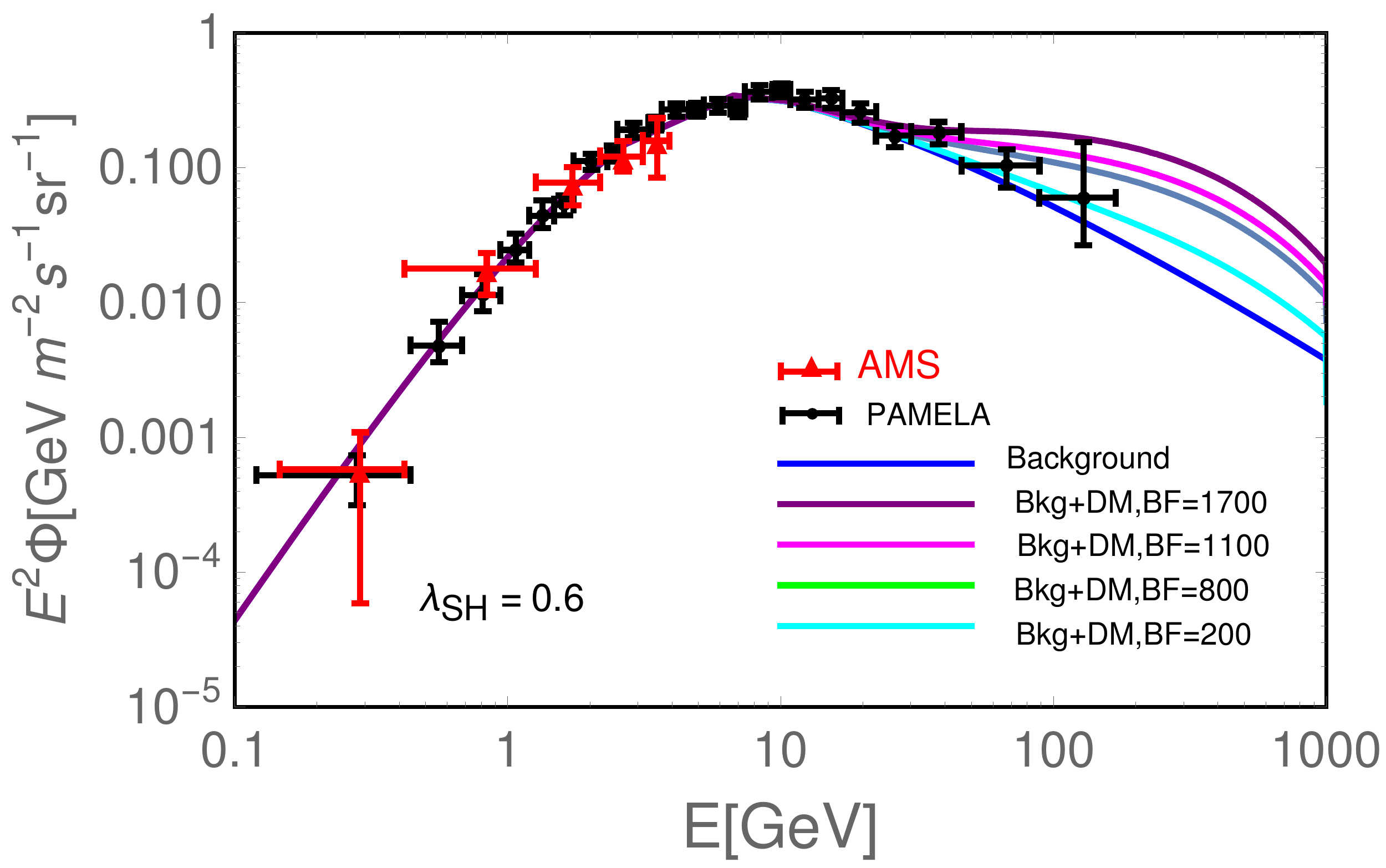}}
\subfigure[]{\includegraphics[height=4.5cm,width=4.9cm]{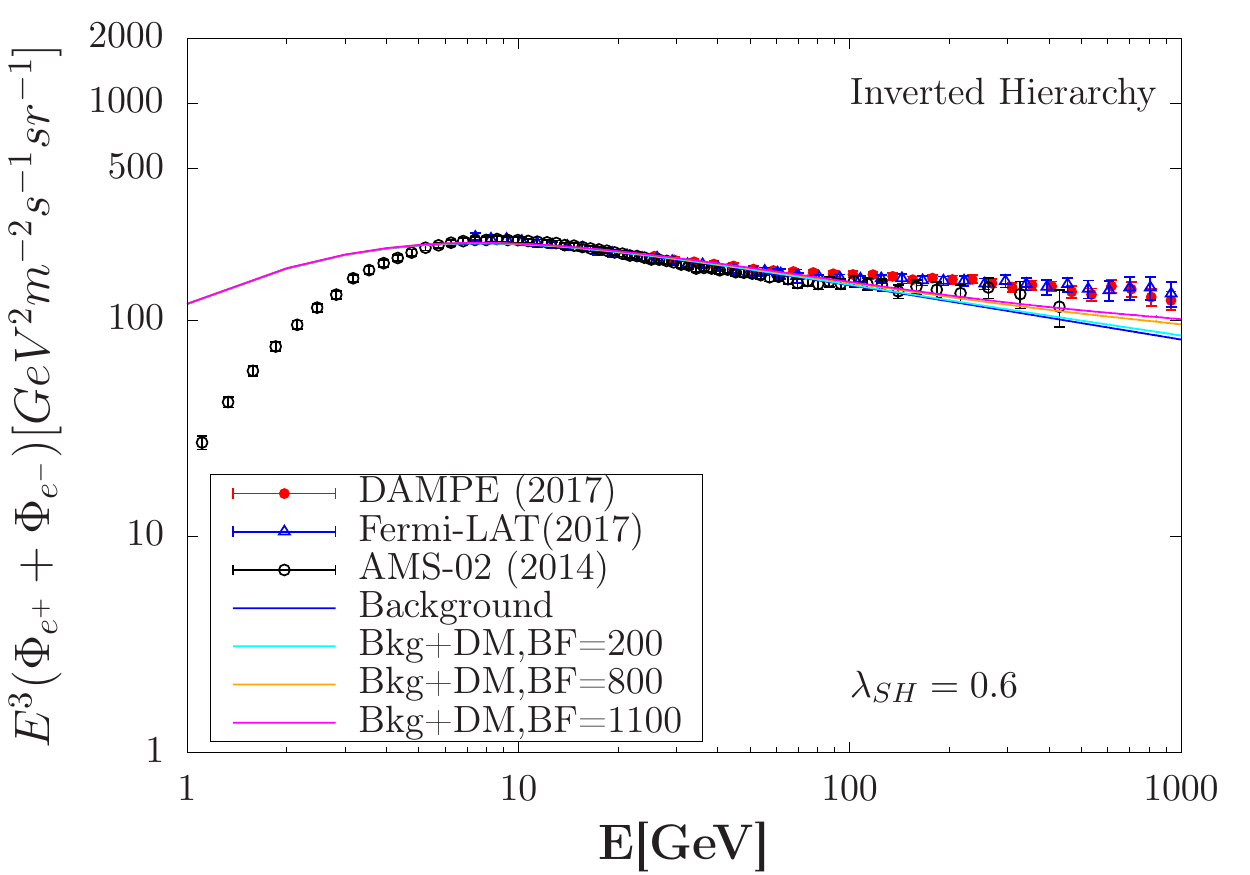}}
\subfigure[]{\includegraphics[height=4.5cm,width=4.9cm]{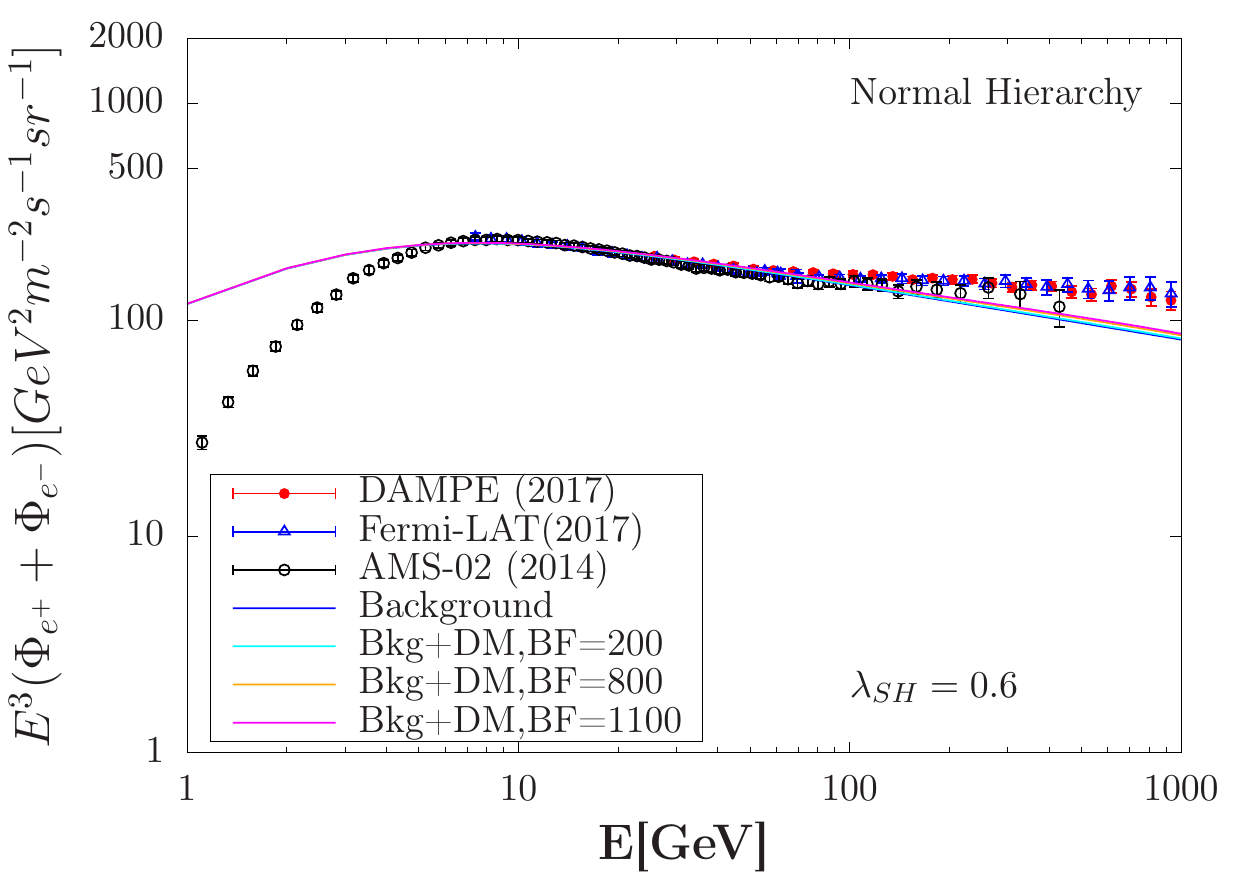}}
\caption{
Same as Figure \ref{result_0.4} but with $\lambda_{SH}=0.6$.} 
\label{result_0.6}
\end{figure}

We present a similar sector of figures in Figure \ref{result_0.6}, the only difference is $\lambda_{SH}=0.6$.
From Figure \ref{result_0.6}(a) we know that the maximum value BF is restricted to less than about 1100, 
otherwise it may lead to an inappropriate antiproton spectrum measured by AMS. Compared with the positron-electron flux, we find that neither IH nor NH scenario can meet the DAMPE and Fermi-LAT measurements. 
They can only meet AMS-02 experiment, and $\lambda_{SH}$ should not be much higher.
In fact, the larger value of $\lambda_{SH}$ is excluded by the AMS and PAMELA experiments, 
as in this case the triplet scalar productions are gradually reduced.

\begin{table}[htbp]
    \centering
    \begin{tabular}{cccccc}
    \hline
    \hline
   $\lambda_{SH}$&0.30&0.40&0.50&0.60&0.70\\
    \hline
    $\lambda_{S\Delta}$&1.04&1.01&0.98&0.95&0.88\\
    \hline
    N1&0\%&1\%&2\%&3\%&4\% \\
    \hline
    N2&94\%&90\%&84\%&76\%&66\% \\
    \hline
    BF&$\lesssim 4200$&$\lesssim 2500$&$\lesssim 1600$&$\lesssim1100$&$\lesssim 800$\\
    \hline
    \hline
    \end{tabular}
    \caption{The fraction scattering cross section $N_i$ and maximum value of BFs at different values of $\lambda_{SH}/\lambda_{S\Delta}$.}
    \label{tabp1}
 \end{table}

We also select some feature points that meet the requirement of relic density, 
direct detection constraints, antiproton flux exhibited by AMS and PAMELA experimental results in Table \ref{tabp1}.
The fraction scattering cross section $N_i$ and the maximum value of BF that allowable are shown, 
corresponding to different values of $\lambda_{SH}/\lambda_{S\Delta}$.
We see that each value of $\lambda_{SH}$ has an upper limit on BF.
As $\lambda_{SH}$ becomes larger, the BF value is reduced correspondingly, while the semi-annihilation contribution enhances, 
though only stands for a small fraction of the total.
And small values of BF and $\lambda_{S\Delta}$ will also lead to bad fit to DAMPE and Fermi-LAT measurements.
After the detailed analysis above, we conclude that
 to meet appropriate antiproton spectrum and electron-positron flux, especially for DAMPE and Fermi-LAT,
we demand the value of $\lambda_{SH}$ not large, i.e., smaller than 0.6.
As a result, the fraction of semi-annihilation cross section is less than $3\%$.

\begin{table}[htpb]
    \centering
    \begin{tabular}{cccc}
    \hline
    \hline
   N1&$\lambda_{SH}$&$\lambda_{S\Delta}$&BF\\
    \hline
    1\%&0.36&1.03&2600 \\
    \hline
    2\% &0.44&1.01&1900\\
    \hline
    3\% &0.56&0.97&1500\\
    \hline
    \hline
    \end{tabular}
    \caption{ Benchmarks for fitting cosmic-ray antiproton flux and electron-positron flux. }
    \label{tabp2}
 \end{table}

\begin{figure}[htbp]
\centering
\subfigure[]{\includegraphics[height=4.5cm,width=4.9cm]{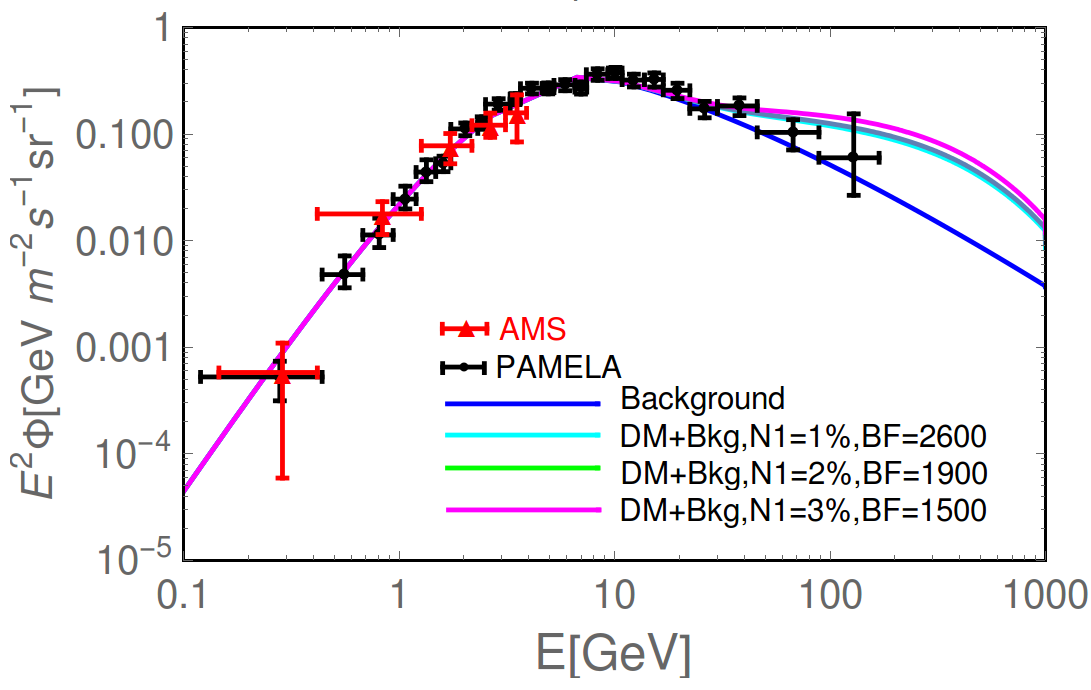}}
\subfigure[]{\includegraphics[height=4.5cm,width=4.9cm]{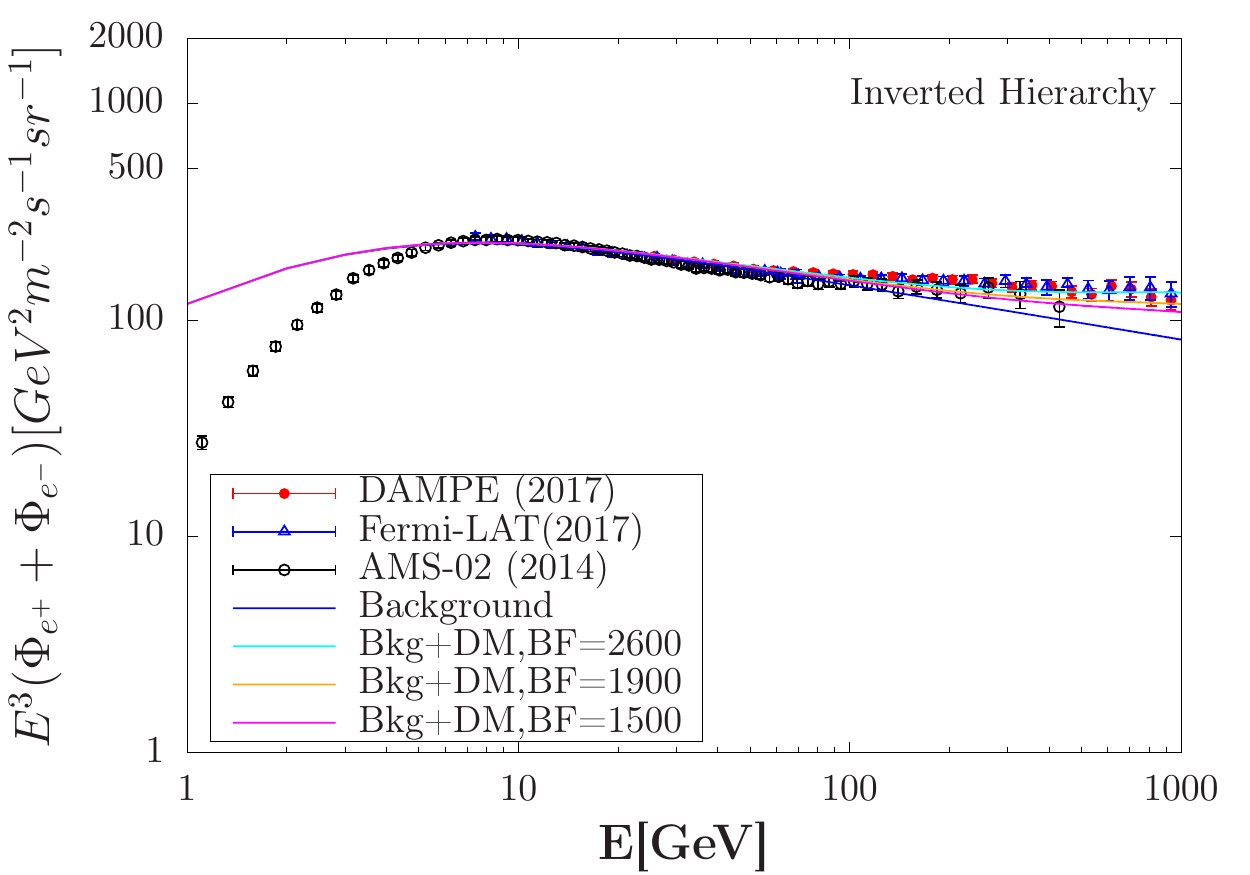}}
\subfigure[]{\includegraphics[height=4.5cm,width=4.9cm]{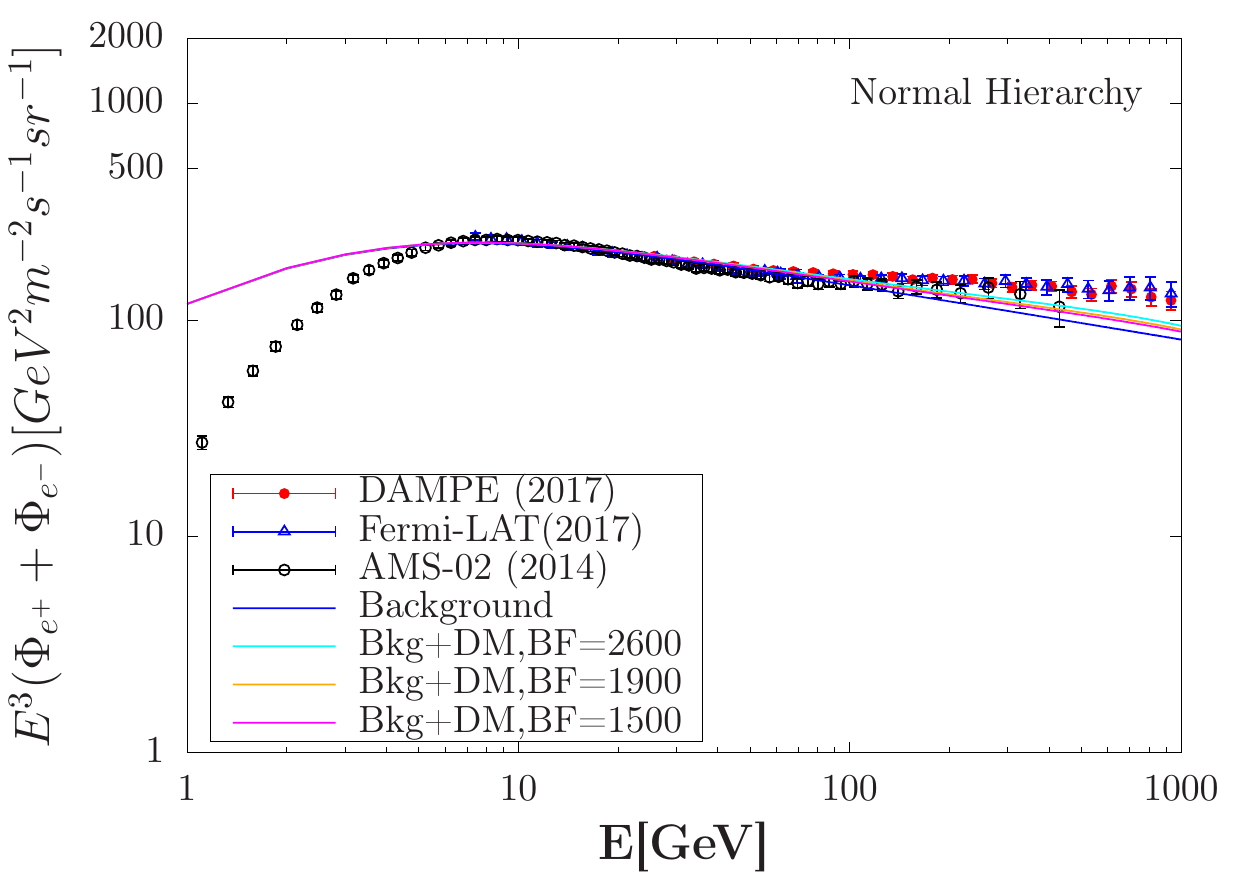}}
\caption{
Background and background + DM (Bkg+DM) of cosmic-ray antiproton flux in (a), 
and electron-positron flux in (b) and (c) with inverted hierarchy (IH) and normal hierarchy (NH) scenario, respectively,
with benchmarks list in Table \ref{tabp2}. %at $\lambda_{SH}$=0.4 and varied BF, 
The data points in (a) labels the PAMELA\cite{Adriani:2010rc} and AMS\cite{Aguilar:2002ad} measurements.
The data points in (b) and (c) stand for the AMS-02\cite{Aguilar:2014mma}, Fermi-LAT\cite{Abdollahi:2017nat} and DAMPE\cite{Ambrosi:2017wek} measurements.
}
\label{bf1}
\end{figure}

To summarize, we give some benchmarks listed in Table \ref{tabp2} which can fit the experiments well with $N_1 <3\%$.
The comparisons with the experiments are presented in Figure \ref{bf1}.

\section{Summary}
 \label{sec:sum}

The idea of relating singlet scalar with $Z_2$ symmetry to Type-II seesaw mechanism
has been extensively studied in the literatures, see for example, Refs.\cite{Dev:2013hka, Dev:2013ff}.
In this paper, we combine the Type-II Seesaw mechanism with a complex singlet scalar of $Z_3$ symmetry
to solve the origin of neutrino mass and dark matter beyond SM in one framework.
The new cubic term $S^3+S^{\dagger3}$ would result in semi-annihilation effects contributing to $\Omega h^2$.
We consider the dark matter in the heavy mass region ($M_S > M_{\Delta}$) and degenerate triplet scalar masses for numerical analysis.
We find the contraints from perturbativity, perturbative unitarity and vacuum stability on $\lambda_{SH}$ and $\lambda_{S\Delta}$ are weak.
The requirement that the ${\cal V}_{(\bcancel{EW}, Z_3)}$ vacuum is the global minimum, or precisely, ${\cal V}_{(\bcancel{EW}, Z_3)} < {\cal V}_{(EW, \bcancel{Z_3})}$,
gives $\mu_3$ a maximum value, which lead to the fact that the semi-annihilate process contribution fraction ($N_1$) has an upper limitation.
We calculate the viable area of $\lambda_{SH}$ and $\lambda_{S\Delta}$ which are consistent with Planck and PandaX-II measurements at given values of $M_{\Delta}$ and $M_{S}$.
As the dark mass $M_{S}$ increases, the semi-annihilation process contribution fraction $N_1$ decreases gradually at the permitte maximum value of $\lambda_{SH}$ and $\mu_3$.

When the triplets and the W, Z boson pairs are produced from DM (co)annihilation,  
they will be responsible for the interpretation of the excess of electron-positron flux and antiproton spectrum with their subsequent decays.
We alleviate the leptophilic properties of the DM to enhance the semi-annihilation effects and calculate those fluxes in both NH and IH scenarios. 
We find that, fitting the antiproton spectrum measured by AMS and PAMELA shows that the maximum value of BF decreases with the increase of $\lambda_{SH}$,
and to fit the DAMPE and Fermi-LAT measurements one requires a large BF and small $\lambda_{SH}$.
To fit simultaneously the electron-positron flux as well as the antiproton spectrum,
a strong restriction is set on the semi-annihilation cross section fraction ($N_1$), for example, for $M_S=3$ TeV, $N_1$ should be less than 3\%.
In either case, the IH scenario is always the favored one.

\acknowledgments

Part of the work is done by the author Aigeng Yang who died of illness and left us far away. 
The authors would like to thank Zhi-Long Han, Chuan-Hung Chen, and Takaaki Nomura for their useful and kind helps during technological implementation. 
Hao Sun is supported by the National Natural Science Foundation of China (Grant No. 12075043, No.12147205).

\appendix
\section{Appendix}
\label{appA}
\subsection{The gauge part}

In Eq.(\ref{kinetic}), the covariant derivative can be expressed as
\begin{equation}
D_{\mu}=\partial_{\mu}+i{\frac{g}{\sqrt{2}}}(T^+W^+_{\mu}+T^-W^-_{\mu})+i\frac{g}{c_{W}}(T^3-s^2_WQ)Z_{\mu}+ieQA_{\mu}
\end{equation}
where $W^{\pm}_{\mu}$, $Z_\mu$ and $A_{\mu}$ are the SM gauge bosons,
$g$ is the gauge coupling of $SU(2)_L$ and $\theta_W$ stands for Weinberg angle with $s_W (c_W)=\sin\theta_W (\cos\theta_W)$ for short.
$Q$ is the charge operator. For the scalar doublet, $T$ are associated to Pauli matrices by
\begin{eqnarray}
T^{\pm} = \frac{1}{2}(\sigma_1\pm \sigma_2),\ \ T^3=\sigma_3
\end{eqnarray}
and $Q={\mathrm{diag}}(1,0)$.
While for the scalar triplet we have $Q={\mathrm{diag}}(2,1,0)$ and $T^{\pm}=T^1 \pm iT^2$ with the $SU(2)_L$ generators chosen as
\begin{equation}
\begin{gathered}
T^1=\frac{1}{\sqrt{2}}\begin{pmatrix}
0&1&0\\1&0&1\\0&1&0
\end{pmatrix},
T^2=\frac{1}{\sqrt{2}}\begin{pmatrix}
0&-i&0\\i&0&-i\\0&i&0
\end{pmatrix},
T^3=\begin{pmatrix}
1&0&0\\0&0&0\\0&0&-1
\end{pmatrix}.
\end{gathered}
\end{equation}
The kinetic terms of the triplet $\Delta$ will change the SM $W^{\pm}$ and $Z$ gauge boson masses,
which can be expressed after spontaneous symmetry breaking as
\begin{equation}
\begin{gathered}
m^2_W=\frac{g^2v^2_0}{4}(1+\frac{2v^2_{\Delta}}{v^2_0}), \ \ m^2_Z=\frac{g^2v^2_0}{4 c^2_W}(1+\frac{4v^2_{\Delta}}{v^2_0}),
\end{gathered}
\end{equation}
therefore affect the SM $\rho$ parameter at the tree level
\begin{equation}
\rho=\frac{m^2_W}{m^2_Z c^2_W}=\frac{1+2v^2_{\Delta}/v^2_0}{1+4v^2_{\Delta}/v^2_0}\ .
\end{equation}
The current electroweak precision data constraints require the $\rho$ parameter to $1.0004^{+0.0003}_{-0.0004}$ \cite{2016Review} and thus give
\begin{eqnarray}
\frac{v_\Delta}{v_0} \lesssim 0.02,\ \ {\mathrm{or}}\ \  v_\Delta \lesssim 5\ GeV,
\end{eqnarray}
One can  find  that the gauge interactions of triplet particles such as $\delta^0W^+W^-$ and $\delta^0ZZ$ vertexes are all proportional to $v_\Delta$. More detailed discussion can be found in Ref.\cite{Chen:2014lla}.
Since in our case we consider the singlet scalar, there is no direct connection between $S$ and gauge fields.

\subsection{The Yukawa part}

The triplet scalar couplings with the leptons through Type-II seesaw mechanism are given, in terms of Yukawa matrix $\textbf{Y}$, by
\begin{eqnarray}
{\cal{L}}^{\mathrm{New}}_{\mathrm{Yukawa}} = - \frac{1}{2} L^{T} {\cal{C}} {\textbf{Y}} i \sigma_2 \Delta L + \mathrm{h.c.}\ ,
\end{eqnarray}
with ${\cal{C}}=i\gamma_0\gamma_2$. $L^T=(\nu^T_\ell,\ell^T)$ is the transpose of the $SU(2)_{L}$ left handed lepton doublet mentioned above.
We can simply obtain the expanded expression
\begin{eqnarray}
{\cal{L}}^{\mathrm{New}}_{\mathrm{Yukawa}} =
- \frac{Y_{ij}}{2} \nu^T_i {\cal{C}} \nu_j \frac{v_{\Delta}+\delta^0+i\eta^0}{\sqrt{2}}
+ Y_{ij} \nu^T_i {\cal{C}} \ell_j \frac{\delta^+}{\sqrt{2}}
+ \frac{Y_{ij}}{2} \ell^T_i {\cal{C}} \ell_j {\delta}^{++} + \mathrm{h.c.}\ ,
\end{eqnarray}
with $i,j=1,2,3$. Therefore $\textbf{Y}$ is related to the Majorana neutrino mass matrix
in flavour eigenstates by $\textbf{m}_{\nu}=v_{\Delta} \textbf{Y} /\sqrt{2}$.
Assume that the physical neutrino mass matrix is $\textbf{m}_{\nu}^{\mathrm{diag}}=\mathrm{diag}(m_1,m_2,m_3)$,
with the help of Pontecorvo-Maki-Nakagawa-Sakata (PMNS) matrix $U_{PMNS}$, the Yukawa matrix can thus be written as
\begin{equation}
\begin{aligned}
{\textbf{Y}}&=\frac{\sqrt{2}}{v_{\Delta}}U^*_{PMNS}\textbf{m}^{\mathrm{diag}}_{\nu} U^{\dagger}_{PMNS}
=\frac{\sqrt{2}}{v_{\Delta}}{\begin{pmatrix}m_{ee}&m_{e\mu}&m_{e\tau}\\m_{e\mu}&m_{\mu\mu}&m_{\mu\tau}\\m_{e\tau}&m_{\mu\tau}&m_{\tau\tau}\end{pmatrix}}.
\end{aligned}
\end{equation}
The $U_{PMNS}$ matrix can be parametrized by
\begin{align} \nonumber
U_{PMNS}=& \left(
\begin{array}{ccc}
c_{12}c_{13}&c_{13}s_{12}&e^{-i\delta}s_{13}\\
-c_{12}s_{13}s_{23}e^{i\delta}-c_{23}s_{12}&c_{12}c_{23}-e^{i\delta}s_{12}s_{13}s_{23}&c_{13}s_{23}\\
s_{12}s_{23}-e^{i\delta}c_{12}c_{23}s_{13}&-c_{23}s_{12}s_{13}e^{i\delta}-c_{12}s_{23}&c_{13}c_{23}
\end{array}
\right) \\
&\times
\mbox{diag}\left(e^{i\Phi_1/2},1,e^{i\Phi_2/2} \right)
\end{align}
with $s_{ij}=\sin\theta_{ij}$, $c_{ij}=\cos\theta_{ij}$ and $\delta$, $\Phi_i$ are the Dirac, Majorana CP phases respectively.
For illustration purposes, we set the two Majorana phases and the lightest neutrino mass to be zero and the masses obey
\begin{itemize}
\item The normal hierarchy (NH) scenario ($m_1<m_2<m_3$) with $m_{2}=(m^2_1+\Delta m^2_{21})^{1/2}$, $m_{3}=(m^2_1+\Delta m^2_{31})^{1/2}$,
\item The inverted hierachy (IH) scenario ($m_3<m_1<m_2$) with $m_1=(m_3^2+\Delta m^2_{31})^{1/2}$, $m_2=(m^2_1+\Delta m^2_{21})^{1/2}$,
\end{itemize}
where $\Delta m^2_{ij}=m^2_i-m^2_j$.
Using the central values of the global analysis based on the neutrino oscillation data \cite{Esteban:2016qun}:
\begin{eqnarray}
\nonumber \Delta m_{21}^2 = 7.50\times10^{-5}{\rm eV}^2\,,&\quad&{|\Delta m_{31}^2|}= 2.524(2.514)\times10^{-3}{\rm eV}^2\,,\nonumber\\
\sin^2\theta_{12}= 0.306\,,&\quad&\sin^2\theta_{23}=0.441(0.587)\,,\nonumber\\
\delta=261^\circ\;(277^\circ)\,,&\quad&\sin^2\theta_{13}=0.02166\;(0.02179)\ ,
\end{eqnarray}
where the values in parenthese correspond to the IH scenario, we obtain the Yukawa coupling matrix
\begin{align}\label{decaymodel1}
{\textbf{Y}}^{\mathrm{NH}}= \frac{10^{-2}{\rm eV}}{v_\Delta} \times
\left(
\begin{array}{ccc}
0.1558+0.0336i&0.2232 - 0.5050i&-0.3432 - 0.5686 i\\0.2232 - 0.5050i& 2.5103-0.0584i&1.0963- 0.03901i\\2.1403-0.0078i&1.0963-0.0390i&3.0004+0.0566i
\end{array}
\right),
\end{align}
and
\begin{align}\label{decaymodel2}
{\textbf{Y}}^{\mathrm{IH}}= \frac{10^{-2}{\rm eV}}{v_\Delta} \times
\left(
\begin{array}{ccc}
3.6631&-1.2660-0.4156i&1.4057-0.3487i\\
-1.2660-0.4158i&0.8693+0.2874i&-1.0963-0.03901i\\
1.4057-0.3487i&-1.0963-0.03901i&1.1869-0.2676i
\end{array}
\right)\ .
\end{align}
With the help of these known Yukawa coupling values, the branching ratios of the triplet scalars
can be calculated and listed in Table \ref{Br} for both the IH and NH scenarios.
The decay of $\delta^{\pm\pm}$ final states produces a fraction of $e:\mu:\tau \approx$ $1:0.2:0.3$ and $1:0.5:1.4$ in IH and NH scenario, respectively.
The electron-rich final states in IH case is pointing to the excess of the spectrum indicated by the AMS-02, DAMPE and Fermi-LAT measurements.
We will give a more detailed discussion in the following section.

\begin{table}[htpb]
    \centering
    \begin{tabular}{ccccccccc}
    \hline
    \hline
    $\delta^{\pm\pm}$  & IH & NH & $\delta^{0}/\eta^{0}$ & IH & NH & $\delta^{\pm}$  & IH& NH\\
    \hline
     Br($e^{\pm}e^{\pm}$)& 51.80\%& 1.32\% & Br($\nu_{e}\nu_{e}$)  &  51.80\% & 1.32\% &Br($e^{\pm}\nu_{e}$)&51.80\%&1.32\%\\
     Br($\mu^{\pm}\mu^{\pm}$)& 3.24\% & 3.28\% & Br($\nu_{\mu}\nu_{\mu}$)  &  3.24\% & 3.28\% &Br($e^{\pm}\nu_{\mu}$)&6.86\%&1.58\%\\
    Br($\tau^{\pm}\tau^{\pm}$)& 5.72\%& 4.68\%& Br($\nu_{\tau}\nu_{\tau}$)  & 5.72\% & 4.68\% &Br($e^{\pm}\nu_{\tau}$)&8.10\%&22.9\%\\
     Br($e^{\pm}\mu^{\pm}$)& 13.70\%& 31.7\%& Br($\nu_{e}\nu_{\mu}$)  & 13.70\% & 31.7\% &Br($\mu^{\pm}\nu_{e}$)&6.86\%&1.58\%\\
     Br($e^{\pm}\tau^{\pm}$)& 16.20\%& 45.9\%& Br($\nu_{e}\nu_{\tau}$)  & 16.20\% & 45.9\% &Br($\mu^{\pm}\nu_{\mu}$)&3.24\%&3.28\%\\
    Br($\mu^{\pm}\tau^{\pm}$)& 9.30\%& 1.25\%& Br($\nu_{\mu}\nu_{\tau}$)  & 9.30 \% & 1.25\% &Br($\mu^{\pm}\nu_{\tau}$)&4.65\%&6.26\%\\
      &&&&&&Br($\tau^{\pm}\nu_{e}$)&8.10\%& 22.9\%\\
         &&&&&&Br($\tau^{\pm}\nu_{\mu}$)&4.65\%&62.6\%\\
        &&&&&&Br($\tau^{\pm}\nu_{\tau}$)&5.72\%&4.68\%\\
     \hline
     \hline
    \end{tabular}
    \caption{Branching radios (Br) of triplet scalars decay calculated by CalcHEP \cite{Belyaev:2012qa} in NH, IH scenarios,
    with the corresponding Yukawa coupling values listed in Eq.(\ref{decaymodel1}) and Eq.(\ref{decaymodel2}).}
\label{Br}
\end{table}

The singly- and doubly-charged scalars in the type-II Seesaw model also contribute to the lepton flavor violating (LFV) processes,
therefore put the stringent lower bounds based on the current experiment \cite{Akeroyd:2009nu, Fukuyama:2009xk}, 
given $v_\Delta \cdot M_\Delta  \geq 150\ {\rm eV}\cdot {\rm GeV}$, where $M_\Delta$ is the triplet mass. 
The limits from LHC direct searches for doubly charged Higgs bosons 
also require $M_{\Delta} \gtrsim 770-870~{\rm GeV}$ \cite{Aaboud:2017qph,ATLAS:2014kca}. 

\section{Formulas}
\label{appB}
The first submatrix ${\cal M}_1$ of the scattering whose initial and final states have charge zero: 
$E_1=($ $G^+\delta^-$, $\delta^+G^-$, $h\eta^0$, $\delta^0G^0$, $G^0\eta^0$, $h\delta^0$, $hs$, $\delta^0S$, $G^0S$, $\eta^0S$, $hS^*$, $\delta^0S^*$, $G^0S^*$, $\eta^0S^*)$.
The first six states have discrete $Z_3$ symmetry with charge $X=0$, and the last eight states have $X=2$. 
One can find:
\begin{equation}
{\scriptsize{\cal M}_1=
\left(
\begin{array}{cccccccccccccc}
\lambda_1+\frac{\lambda_4}{2}&0&\frac{i\lambda_4}{2\sqrt{2}}&-\frac{i\lambda_4}{2\sqrt{2}}&\frac{\lambda_4}{2\sqrt{2}}&\frac{\lambda_4}{2\sqrt{2}}&0&0&0&0&0&0&0&0\\
0&\lambda_1+\frac{\lambda_4}{2}&-\frac{i\lambda_4}{2\sqrt{2}}&\frac{i\lambda_4}{2\sqrt{2}}&\frac{\lambda_4}{2\sqrt{2}}&\frac{i\lambda_4}{2\sqrt{2}}&0&0&0&0&0&0&0&0\\
\frac{i\lambda_4}{2\sqrt{2}}&-\frac{i\lambda_4}{2\sqrt{2}}&(\lambda_1+\lambda_4)&0&0&0&0&0&0&0&0&0&0&0\\
-\frac{i\lambda_4}{2\sqrt{2}}&\frac{i\lambda_4}{2\sqrt{2}}&0&\lambda_1+\lambda_4&0&0&0&0&0&0&0&0&0&0\\
\frac{\lambda_4}{2\sqrt{2}}&\frac{\lambda_4}{2\sqrt{2}}&0&0&\lambda_1+\lambda_4&0&0&0&0&0&0&0&0&0\\
\frac{\lambda_4}{2\sqrt{2}}&\frac{\lambda_4}{2\sqrt{2}}&0&0&0&\lambda_1+\lambda_4&0&0&0&0&0&0&0&0\\
0&0&0&0&0&0&\lambda_{SH}&0&0&0&0&0&0\\
0&0&0&0&0&0&0&\lambda_{S\Delta}&0&0&0&0&0&0\\
0&0&0&0&0&0&0&0&\lambda_{SH}&0&0&0&0&0\\0&0&0&0&0&0&0&0&0&\lambda_{S\Delta}&0&0&0&0\\
0&0&0&0&0&0&0&0&0&0&\lambda_{SH}&0&0&0\\0&0&0&0&0&0&0&0&0&0&0&\lambda_{S\Delta}&0&0\\
0&0&0&0&0&0&0&0&0&0&0&0&\lambda_{SH}&0\\
0&0&0&0&0&0&0&0&0&0&0&0&0&\lambda_{S\Delta}
\end{array}
\right).}
\end{equation}
\\
The second submatrix ${\cal M}_2$ of the scattering whose initial and final states have charge zero:
$E_2=($ $G^+G^-$, $\delta^+ \delta^-$, $\frac{G^0 G^0}{\sqrt{2}}$, $\frac{\eta^0 \eta^0}{\sqrt{2}}$,
$\frac{hh}{2}$, $\frac{\delta^0 \delta^0}{\sqrt{2}}$, $\delta^{++} \delta^{--}$, $SS^{*})$.
All these states have discrete $Z_3$ charge $X=0$. One can find:
\begin{equation}
{\scriptsize{\cal M}_2=
\left(
\begin{array}{cccccccc}
4\lambda & \lambda_1+\frac{\lambda_4}{2}&\frac{4\lambda}{2\sqrt{2}}&\frac{\lambda_1}{\sqrt{2}}&\frac{4\lambda}{2\sqrt{2}}&\frac{\lambda_1}{\sqrt{2}}&\lambda_1+\lambda_4&\lambda_{SH}\\
\lambda_1+\frac{\lambda_4}{2}&4\lambda_2+2\lambda_3&\frac{2\lambda_1+\lambda_4}{2\sqrt{2}}&\sqrt{2}(\lambda_2+\lambda_3)&\frac{2\lambda_1+\lambda_4}{2\sqrt{2}}&\sqrt{2}(\lambda_2+\lambda_3)&2(\lambda_2+\lambda_3)&\lambda_{S\Delta}\\
\frac{4\lambda}{2\sqrt{2}} &\frac{2\lambda_1+\lambda_4}{2\sqrt{2}}&3\lambda&\frac{\lambda_1+\lambda_4}{2}&\lambda&\frac{\lambda_1+\lambda_4}{2}&\frac{\lambda_1}{\sqrt{2}}&\frac{\lambda_{SH}}{\sqrt{2}}\\
\frac{\lambda_1}{\sqrt{2}}&\sqrt{2}(\lambda_2+\lambda_3)&\frac{\lambda_1+\lambda_4}{2}&3(\lambda_2+\lambda_3)&\frac{\lambda_1+\lambda_4}{2}&(\lambda_2+\lambda_3)&\sqrt{2}\lambda_2&\frac{\lambda_{S\Delta}}{\sqrt{2}}\\
\frac{4\lambda}{2\sqrt{2}}&\frac{2\lambda_1+\lambda_4}{2\sqrt{2}}&\lambda&\frac{\lambda_1+\lambda_4}{2}&3\lambda&\frac{\lambda_1+\lambda_4}{2}&\frac{\lambda_1}{\sqrt{2}}&\frac{\lambda_{SH}}{\sqrt{2}}\\
\frac{\lambda_1}{\sqrt{2}}&\sqrt{2}(\lambda_2+\lambda_3)&\frac{\lambda_1+\lambda_4}{2}&(\lambda_2+\lambda_3)&\frac{\lambda_1+\lambda_4}{2}&3(\lambda_2+\lambda_3)&\sqrt{2}\lambda_2&\frac{\lambda_{S\Delta}}{\sqrt{2}}\\
(\lambda_1+\lambda_4)&2(\lambda_2+\lambda_3)&\frac{\lambda_1}{\sqrt{2}}&\sqrt{2}\lambda_2&\frac{\lambda_1}{\sqrt{2}}&\sqrt{2}\lambda_2&4(\lambda_2+\lambda_3)&\lambda_{S\Delta}\\
\lambda_{SH}&\lambda_{S\Delta}&\frac{\lambda_{SH}}{\sqrt{2}}&\frac{\lambda_{S\Delta}}{\sqrt{2}}&\frac{\lambda_{SH}}{\sqrt{2}}&\frac{\lambda_{S\Delta}}{\sqrt{2}}&\lambda_{S\Delta}&4\lambda_{S}
\end{array}
\right).}
\end{equation}
\\
The third submatrix ${\cal M}_3$ of the scattering whose initial and final states have charge zero:
$E_3=($ $hG^0$, $\delta^0\eta^0$, $\frac{SS}{\sqrt{2}}$, $\frac{S^*S^*}{\sqrt{2}})$.
The first two states have discrete $Z_3$ charge $X=0$, the states $\frac{S^*S^*}{\sqrt{2}}$ and $\frac{SS}{\sqrt{2}}$ have $X=2$ and $X=1$ respectively. 
One can find:
\begin{equation}
{\scriptsize{\cal M}_3=\left(
\begin{array}{cccc}
2\lambda&0&0&0\\
0&2(\lambda_2+\lambda_3)&0&0\\
0&0&\lambda_S&0\\
0&0&0&\lambda_S
\end{array}
\right).}
\end{equation}
\\
The fourth submatrix ${\cal M}_4$ of the scattering whose initial and final states are charge one:
$E_4=($ $hG^+$, $\delta^0G^+$, $G^0G^+$, $\eta^0G^+$, $h\delta^+$, $\delta^0\delta^+$, $G^0\delta^+$,
$\eta^0\delta^+$, $\delta^{++}\delta^-$, $\delta^{++}G^-$, $SG^+$, $S\delta^+$, $S^*G^+$, $S^*\delta^+)$.
The first ten states have discrete $Z_3$ charge $X=0$, the states $SG^+, S\delta^+$ and $S^*G^+, S^*\delta^+$ have $X=1$ and $X=2$ respectively.
One can find:
\begin{equation}
{\tiny{\cal M}_4=\left(
\begin{array}{cccccccccccccc}
2\lambda&0&0&0&0&\frac{\lambda_4}{2\sqrt{2}}&0&-i\frac{\lambda_4}{2\sqrt{2}}&-\frac{\lambda_4}{2}&0&0&0&0&0\\
0&\lambda_1&0&0&\frac{\lambda_4}{2\sqrt{2}}&0&i\frac{\lambda_4}{2\sqrt{2}}&0&0&0&0&0&0&0\\
0&0&2\lambda&0&0&i\frac{\lambda_4}{2\sqrt{2}}&0&\frac{\lambda_4}{2\sqrt{2}}&-i\frac{\lambda_4}{2}&0&0&0&0&0\\
0&0&0&\lambda_1&-i\frac{\lambda_4}{2\sqrt{2}}&0&\frac{\lambda_4}{2\sqrt{2}}&0&0&0&0&0&0&0\\
0&\frac{\lambda_4}{2\sqrt{2}}&0&-\frac{\lambda_4}{2\sqrt{2}}&\frac{2\lambda_1+\lambda_4}{2}&0&0&0&0&-\frac{\lambda_4}{2}&0&0&0&0\\
\frac{\lambda_4}{2\sqrt{2}}&0&i\frac{\lambda_4}{2\sqrt{2}}&0&0&2(\lambda_2+\lambda_3)&0&0&-\sqrt{2}\lambda_3&0&0&0&0&0\\
0&i\frac{\lambda_4}{2\sqrt{2}}&0\frac{\lambda_4}{2\sqrt{2}}&0&0&0&\frac{2\lambda_1+\lambda_4}{2}
&0&0&-i\frac{\lambda_4}{2}&0&0&0&0\\
-i\frac{\lambda_4}{2\sqrt{2}}&0&\frac{\lambda_4}{2\sqrt{2}}&0&0&0&0&2(\lambda_2+\lambda_3)&-i\sqrt{2}\lambda_3&0&0&0&0&0\\
-\frac{\lambda_4}{2}&0&-i\frac{\lambda_4}{2}&0&0&-\sqrt{2}\lambda_3&o&-i\sqrt{2}\lambda_3&2(\lambda_2+\lambda_3)&0&0&0&0&0\\
0&0&0&0&-\frac{\lambda_4}{2}&0&-i\frac{\lambda_4}{2}&0&0&\lambda_1+\lambda_4&0&0&0&0\\
0&0&0&0&0&0&0&0&0&0&\lambda_{SH}&0&0&0\\
0&0&0&0&0&0&0&0&0&0&0&\lambda_{S\Delta}&0&0\\
0&0&0&0&0&0&0&0&0&0&0&0&\lambda_{SH}&0\\
0&0&0&0&0&0&0&0&0&0&0&0&0&\lambda_{S\Delta}
\end{array}
\right).}
\end{equation}
\\
The fifth submatrix $M_5$ of the scattering whose initial and final states have charge two:
$E_5=($ $\frac{G^+G^+}{\sqrt{2}}$, $\frac{\delta^+\delta^+}{\sqrt{2}}$, $\delta^+G^+$, $\delta^{++}\delta^0$, $\delta^{++}\eta^0$,
$\delta^{++}G^0$, $\delta^{++}h$, $\delta^{++}S$, $\delta^{++}S^{*})$.
The first seven states have discrete $Z_3$ charge $X=0$, the states $\delta^{++}S$ and $\delta^{++}S^{*}$ have $X=1$ and $X=2$ respectively. 
One can find:
\begin{equation}
{\scriptsize{\cal M}_5=
\left(
\begin{array}{ccccccccc}
2\lambda&0&0&0&0&0&0&0&0\\
0&2\lambda_2+\lambda_3&0&-\lambda_3&-i\lambda_3&0&0&0&0\\
0&0&\frac{2\lambda_1+\lambda_4}{2}&0&0&-i\frac{\lambda_4}{2}&-\frac{\lambda_4}{2}&0&0\\
0&-\lambda_3&0&2\lambda_2&0&0&0&0&0\\
0&-i\lambda_3&0&0&2\lambda_2&0&0&0&0\\
0&0&-i\frac{\lambda_4}{2}&0&0&\lambda_1&0&0&0\\
0&0&-\frac{\lambda_4}{2}&0&0&0&\lambda_1&0&0\\
0&0&0&0&0&0&0&\lambda_{S\Delta}&0\\
0&0&0&0&0&0&0&0&\lambda_{SH}
\end{array}
\right).}
\end{equation}
\\
The sixth submatrix ${\cal M}_6$ of the scattering whose initial and final states have charge three: $E_6=($ $\delta^{++}G^+$, $\delta^{++}\delta^+)$. 
All states have discrete $Z_3$ charge $X=0$. One can find:
\begin{equation}
{\scriptsize{\cal M}_6=
\left(
\begin{array}{cc}
\lambda_1+\lambda_4&0\\
0&2(\lambda_2+\lambda_3)
\end{array}
\right).}
\end{equation}
\\
The last submatrix ${\cal M}_7$ of the scattering whose initial and final states have charge four:
$E_7=\frac{\delta^{++}\delta^{++}}{\sqrt{2}}$, its discrete $Z_3$ charge $X=0$. 
The ${\cal M}_7$ is $2(\lambda_2+\lambda_3)$.

The eigenvalues $e^j_i$ of the submatrix ${\cal M}_i$ can be writen as
\begin{equation}
\begin{aligned}
%&e^1_1=\lambda_1+\lambda_4 \leq 8\pi,
&e^1_1=\lambda_1+\lambda_4,
e^2_1=\lambda_1,
e^3_1=\lambda_1+\frac{3}{2}\lambda_4,
%e^4_1=\lambda_{SH} \leq 8 \pi,\\
%&e^5_1=\lambda_{S\Delta} \leq 8 \pi,
e^4_1=\lambda_{SH},\\
&e^5_1=\lambda_{S\Delta} ,
e^1_2=2\lambda,
e^2_2=2\lambda_2,
e^3_2=2(\lambda_2+\lambda_3), \\
&e^4_2=\lambda+\lambda_2+2\lambda_3+\sqrt{\lambda^2-2\lambda\lambda_2+\lambda^2_2-4\lambda\lambda_3+4\lambda_2\lambda_3+4\lambda^2_3+\lambda^2_4}    ,\\
&e^5_2=\lambda+\lambda_2+2\lambda_3-\sqrt{\lambda^2-2\lambda\lambda_2+\lambda^2_2-4\lambda\lambda_3+4\lambda_2\lambda_3+4\lambda^2_3+\lambda^2_4}    ,\\
&e^6_2=\frac{1}{4}Root[A],
e^1_3=2\lambda,
e^2_3=2(\lambda_2+\lambda_3),
%e^3_3=\lambda_S \leq  8\pi,
e^3_3=\lambda_S ,
e^1_4=\lambda_1+\lambda_4,
e^2_4=\lambda_1,\\
%&e^3_4=\frac{2\lambda_1+3\lambda_4}{2} ,
&e^3_4=\lambda_1+\frac{3\lambda_4}{2},
e^4_4=2\lambda,
e^5_4=2\lambda_2,
e^6_4=2(\lambda_2+\lambda_3),
e^7_4=\lambda_1-\frac{\lambda_4}{2}, \\
&e^8_4=\frac{1}{4}[4\lambda+4\lambda_2+8\lambda_3+\sqrt{(4\lambda-4\lambda_2-8\lambda_3)^2+16\lambda^2_4}],\\
&e^9_4=\frac{1}{4}[4\lambda+4\lambda_2+8\lambda_3-\sqrt{(4\lambda-4\lambda_2-8\lambda_3)^2+16\lambda^2_4}], \\
&e^{10}_4=\lambda_{SH},
e^{11}_4=\lambda_{S\Delta},
e^1_5=\lambda_1+\lambda_4 ,
e^2_5=\lambda_1,
e^3_5=2\lambda,
e^4_5=2\lambda_2,
e^5_5=2(\lambda_2+\lambda_3),\\
&e^6_5=\lambda_1-\frac{\lambda_4}{2},
e^7_5=2\lambda_2 -\lambda_3,
e^8_5=\lambda_{S\Delta},
e^1_6=\lambda_1+\lambda_4,
e^2_6=2(\lambda_2+\lambda_3),
e^1_7=2(\lambda_2+\lambda_3).
\end{aligned}
\end{equation}
Where we have ignored duplicate eigenvalues for ${\cal M}_i$. The symbol $Root[A]$ stand for the roots of cubic equation,
we will apply Sanmuelson's inequality \cite{1968How} to place restrictions on the region of roots.
\begin{equation}
\begin{aligned}
x^3&+x^2(-24\lambda-32\lambda_2-24\lambda_3-16\lambda_S)+x(-96\lambda^2_1+768\lambda\lambda_2+576\lambda\lambda_3-100\lambda_1\lambda_4-26\lambda^2_4) \\
&+1536\lambda^2_1\lambda_S-12288\lambda\lambda_2\lambda_S-9216\lambda\lambda_3\lambda_S+1600\lambda_1\lambda_4\lambda_S+416\lambda^2_4\lambda_S+1152\lambda\lambda^2_{S\Delta} \\
&-768\lambda_1\lambda_{S\Delta}\lambda_{SH}-400\lambda_4\lambda_{S\Delta}\lambda_{SH}+1024\lambda_2\lambda^2_{SH}+768\lambda_3\lambda^2_{SH}=0 \ .
\end{aligned}
\end{equation}
\bibliography {y0314}
\end{document}